\documentclass[letterpaper,12pt]{article}

\usepackage{endnotes}
\usepackage{amssymb}
\usepackage{amsfonts}
\usepackage{amsmath}
\usepackage{amsthm}
\usepackage{graphicx,color}
\usepackage{xcolor}
\usepackage{comment}
\usepackage{array}

\usepackage{url}

\usepackage{setspace}
\def\R{\mathbb{R}}
\def\endproof{\hfill\diamondsuit}
\def\sH{{\mathcal H}}
\def\sF{{\mathcal F}}

\def\sC{{\mathcal C}}

\def\E{\mathbb{E}}

\def\sF{\mathcal{F}}
\def\P{\mathbb{P}}

\def\N{\mathbb N}

\newcommand{\pld}{\varphi_{\lambda\downarrow}}
\newcommand{\plu}{\varphi_{\lambda\uparrow}}

\newcommand{\rar}{\rightarrow}
\numberwithin{equation}{section}
\theoremstyle{plain}                % title and  number in bold, text italic
\newtheorem{theorem}{Theorem}[section]
\newtheorem{lemma}[theorem]{Lemma}
\newtheorem{proposition}[theorem]{Proposition}

\theoremstyle{definition}           % title and number in bold, text normal
\newtheorem{definition}[theorem]{Definition}
\newtheorem{example}[theorem]{Example}

\newtheorem{assumption}[theorem]{Assumption}

\theoremstyle{remark}               % title and number in italic, text normal

\usepackage[margin=1.25in]{geometry}

\usepackage{ulem} 

\begin{document}

\begin{center}
\large{\bf Uniqueness in Cauchy problems for diffusive real-valued  strict local martingales}$^\ast$\makeatletter{\renewcommand*{\@makefnmark}{}\footnotetext{\hspace{-.35in} $^\ast${The authors have benefited from helpful comments from {an anonymous referee,} Erik Ekstr\"om, Ioannis Karatzas, Martin Larsson, Dan Ocone, Li-Cheng Tsai, Johan Tysk, Kim Weston, {and participants at the Temple/UPenn probability seminar and Intech meetings}. In particular, many thanks to Johannes Ruf for his many valuable suggestions. The second author has been supported by the National Science Foundation under Grant No. DMS 1812679 (2018 - 2022). Any opinions, findings, and conclusions or recommendations expressed in this material are those of the author(s) and do not necessarily reflect the views of the National Science Foundation (NSF). The corresponding author is Kasper Larsen.  Umut \c{C}etin has email: u.cetin@lse.ac.uk and Kasper Larsen has email: KL756@math.rutgers.edu.}\makeatother}}
\end{center}

 %We should like to thank blah blah Hao Xing, Steve Shreve, the anonymous referee, the anonymous Associate Editor, and the Co-Editor Pierre Collin-Dufresne for their constructive comments.}

\begin{center}

{ \bf Umut \c{C}et\.in}\\ 
London School of Economics

\ \\

{ \bf Kasper Larsen}\\
Rutgers University
\ \\ \ \\

{\normalsize \today }
\ \\ \ \\

\end{center}

\begin{verse}
{\sc Abstract}: For a  real-valued one dimensional diffusive strict local martingale,, we provide a set of smooth functions in which the Cauchy problem has a unique classical solution  { under a local H\"older condition. Under the weaker Engelbert-Schmidt conditions, we provide a set in which the Cauchy problem has a unique weak solution}. We  exemplify our results using quadratic normal volatility models and the two dimensional Bessel process.

\end{verse}
\ \\
\begin{verse}
{\sc MSC2020 subject classifications:} Primary 60G44. Secondary 60J60.
\end{verse}
\ \\
\begin{verse}

\end{verse}

\newpage

\section{Introduction}
Consider a unique weak solution $\big(\P^x\big)_{x\in\R}$ of the time-homogenous diffusion
\begin{align}\label{dX22intro}
dX_t = \sigma(X_t)dB_t, \quad X_0=x,\quad x\in\R,
\end{align}
where { $B=(B_t)_{t\ge0}$ is a standard one-dimensional Brownian motion, $\sigma:\R\to\R$ is a Borel function, $\sigma(x)^2>0$ for all $x\in\R$, and $\sigma$ satisfies the necessary and sufficient {\em Engelbert-Schmidt conditions} for the existence and uniqueness of a weak solution of \eqref{dX22intro}. } For given continuous data $H:\R\to\R$, consider the Cauchy problem 
\begin{align}\label{CauchyI1}
	\begin{cases}
		&h(0,x) = H(x),\quad x\in \R,\\
		&h_t(t,x) =\frac12\sigma(x)^2h_{xx}(t,x),\quad t>0,\quad x\in\R.
	\end{cases}
\end{align}
 { A function $h:[0,\infty)\times \R\to \R$ is said to be a {\em classical} solution of \eqref{CauchyI1} if $h\in \mathcal{C}^{1,2}$ and satisfies  \eqref{CauchyI1}. On the other hand, $h\in \sC$ is said to be a {\em weak} solution of \eqref{CauchyI1}  if $h(0,x)=H(x)$, $x\in\R$,  and 
		\begin{align} \label{e:weakInt}
			\int_0^{\infty} \int_{-\infty}^{\infty}\Big(\frac{2}{\sigma^2(x)}f_t(t,x)+f_{xx}(t,x)\Big)h(t,x)dxdt=0,
		\end{align}
		for all $f:(0,\infty)\times \R\to \R$ in $\sC_c^{1,2}$  (see Appendix \ref{s:weak} for more on weak solutions).
}

For $\sigma$ in \eqref{dX22intro}  with at most linear growth, the solution  of \eqref{dX22intro} becomes a martingale.  In that case, when the data $H$ has at most polynomial growth, there exists at most one classical solution of  \eqref{CauchyI1} of polynomial growth (see Theorem 5.7.6 in Karatzas and Shreve \cite{KS88} for when $\sigma$ is  {additionally continuous). When $\sigma$ in \eqref{dX22intro} satisfies  the Engelbert-Schmidt conditions and ensures $X_t$  is a positive martingale, Bayraktar and   Xing \cite{BH2010} prove uniqueness in the set of at most linearly growing solutions of  \eqref{CauchyI1} for data $H$ of at most linear growth. When  $\sigma$  is locally $\frac12$-H\"older continuous and of at most linear growth and $H$ is of at most polynomial growth, the function 
\begin{align}\label{def_h}
\begin{split}
h^*(t,x):&=\E^x[H(X_t)],\quad t\ge0, \quad x\in\R,
\end{split}
\end{align}
is the unique classical solution of \eqref{CauchyI1} of at most polynomial growth (see Theorem 6.1 in Janson and Tysk \cite{JT2006}). }

When the solution of \eqref{dX22intro} is a strict local martingale, the data  $H(\xi):=\xi$ gives an example where uniqueness of \eqref{CauchyI1} fails because both $h^*$ defined in \eqref{def_h} and $h^\circ(t,x):=x$ are of at most linear growth and both $h^*$ and $h^\circ$  solve \eqref{CauchyI1}.\footnote{ This lack of uniqueness produces several complications. For example, Monte-Carlo simulation cannot approximate $h^*$  in \eqref{def_h} because all discretization schemes produce martingales.}  Given a nonnegative strict local martingale with  dynamics \eqref{dX22intro} and $H$ of at most strict sublinear asymptotic growth as $\xi\uparrow\infty$, Theorem 4.3 in Ekstr\"om and Tysk \cite{ET2009} ensures that \eqref{def_h} is the unique classical solution of \eqref{CauchyI1} in the class of strictly sublinearly growing functions under a local $\frac12$-H\"older continuity hypothesis on $\sigma$. More recently, Theorem 6.2 in \c{C}etin \cite{Cetin2018} allows  $H$ to be of at most linear growth and proves uniqueness in the class of strictly sublinearly growning classical solutions of \eqref{CauchyI1} when the solution of \eqref{dX22intro} is nonnegative.\footnote{The existence and uniqueness results in \cite{ET2009} and \cite{Cetin2018} readily extend to the case where $X_t$ takes values in $(-\infty,a)$ and $(a,\infty)$ for $a\in\mathbb{R}$. For example, when $X_t\in (a,\infty)$, we can consider $\tilde{X}_t := X_t-a$ valued in $(0,\infty)$ with dynamics
	$$
	d\tilde X_t = \sigma(\tilde X_t+a)dB_t.
	$$
	Similarly, when $X_t\in (-\infty,a)$, we can consider $\tilde{X}_t := a-X_t$ valued in $(0,\infty)$ with dynamics
	$$
	d\tilde X_t = \sigma(a-\tilde X_t)dB_t.
	$$}

{Our contributions are: Under the standard Engelbert-Schmidt conditions on $\sigma$,  we show that $h^*$ in \eqref{def_h} is the unique weak solution of  \eqref{CauchyI1} in a suitable set $\sH_\lambda$ of continuous functions    satisfying non-trivial growth conditions as $x\to \pm \infty$ provided the data $H(x)$  satisfies certain growth conditions as  $x\to\pm \infty$ (with $H(\xi):=\xi$ allowed).  Under a stronger local $\frac12$-H\"older-continuity condition on $\sigma$ ensuring a unique strong solution to \eqref{dX22intro}, we  prove that $h^*$ in \eqref{def_h} is the unique classical solution of  \eqref{CauchyI1} in  $\sH_\lambda$. Finally, in the last section, also under the local $\frac12$-H\"older-continuity condition on $\sigma$, we provide a martingale regularization technique and use it to show uniqueness when $H(\xi)$ has polynomial growth (provided an exogenous moment condition on the unique strong solution to \eqref{dX22intro})}.

{Our  uniqueness results complement \cite{ET2009}  and \cite{Cetin2018} in the sense that $X_t\in \R$ in \eqref{dX22intro} whereas both \cite{ET2009} and  \cite{Cetin2018} consider $X_t\in[0,\infty)$.  The uniqueness result in \cite{Cetin2018} is proved by a change-of-measure argument, in which the local martingale  $X_t$ acts as the Radon-Nikodym derivative process. However, because $X_t\in\R$,  $X_t$ cannot be used as a Radon-Nikodym derivative and, consequently, the change-of-measure technique of \cite{Cetin2018} is inapplicable. Nevertheless, our proof of uniqueness is similar in spirit and relies on a Radon-Nikodym derivative process based on a $\lambda$-harmonic function $\Phi_{\lambda}$ for a constant $\lambda>0$. Thus, our proof can be seen as an extension of \cite{Cetin2018}, where the change of measure in \cite{Cetin2018} uses the harmonic function $\Phi_0(\xi):=\xi$ for $\xi\geq 0$. Because there is no non-constant harmonic function associated with \eqref{dX22intro}, we use $\lambda$-harmonic functions with $\lambda>0$ (these always exist). Theorem 2.2 in Urusov and Zervos \cite{UZ17} links the asymptotic behavior of minimal $\lambda$-harmonic functions ($\xi\to\pm\infty$) to whether the solution of \eqref{dX22intro} is a true martingale or not. Identifying this asymptotic  behavior when $X_t$ is a strict local martingale was essential in constructing the set of functions $\sH_\lambda$ in which we prove uniqueness.

{ While Feynman-Kac representations have a long history in stochastic analysis, our problem motivation comes from derivatives pricing in financial economics.\footnote{ In financial economics, $h^*$ in \eqref{def_h} serves as an arbitrage-free derivatives price. When $X_t$ in   \eqref{dX22intro}  is a strict local martingale, several arbitrage-free derivatives prices are available. For example,  Cox and Hobson \cite{CH2005}  call $h^*$ in \eqref{def_h} the ``fair" derivatives price whereas Andersen \cite{Andersen2011} calls $h^*$ the ``minimal" derivatives price.} } 
  By allowing $X_t\in \R$ in  \eqref{dX22intro},  we can consider} two widely used strict local martingale models:   Quadratic normal volatility models (often used in financial economics to model stock bubbles, see, e.g., Z\"uhlsdorff \cite{Zuhl2001},  Andersen \cite{Andersen2011}, and Carr, Fisher, and Ruf \cite{CFR2013}) and the logarithm of the two dimensional Bessel process. The class of data functions $H$ covered by our result is given in terms of $\lambda$-harmonic functions and while our class always includes $H$ of at most linear growth, our class typically also  includes  faster growing data $H$. For example, for the  logarithm of the two dimensional Bessel example mentioned above, $H(\xi)$ can grow superlinearly as $\xi\uparrow\infty$.

The many applications of strict local martingales for modeling purposes in financial economics  include: (i) Basak and Cuoco \cite{BC1998}, Hugonnier \cite{Hug2012}, and Chabakauri \cite{Chab2015} show that strict local martingales can appear endogenously in equilibrium theory.  (ii) Cox and Hobson \cite{CH2005}, Jarrow, Protter, and Shimbo \cite{JPS2006}, Heston, Loewenstein,  and Willard \cite{HLW07}, and Andersen \cite{Andersen2011} use strict local martingales for derivatives pricing. (iii) Stochastic portfolio theory as surveyed in  Fernholz and Karatzas \cite{FK2009} uses strict local martingales to model relative arbitrage opportunities. (iv) Karatzas, Lehoczky, Shreve, and Xu \cite{KLSX91}, Kramkov and Schachermayer \cite{KS99}, and Lowenstein and Willard \cite{LW00} exemplify that strict local martingales can appear as dual utility maximizers. More recent references based on nonnegative local martingales include Kardaras, Kreher, and Nikeghbali \cite{KKN2015}, Kramkov and Weston \cite{KW2016}, and Kardaras and Ruf \cite{KR19, KR2020}.  Hulley and Platen \cite{HP2011} and Hulley and Ruf \cite{HR2019} are recent  references based on real-valued local martingales.

Finally, we mention that an alternative to explicitly pin down growth conditions as $\xi\to\pm \infty$  as we do in this paper,  $h$ in \eqref{def_h} can also be uniquely characterized via a smallness property (for details, see the references and results in Section 5.1 in Karatzas and Ruf \cite{KR16}). From a numerical perspective, the smallness characterization can be difficult to implement (see \cite{ELST2010}) whereas our growth conditions as $\xi\to\pm \infty$ are  compatible with standard numerical procedures such as finite difference methods. For example, when the solution of \eqref{dX22intro} is a strict local martingale, the solution of the Cauchy problem \eqref{CauchyI1} can exhibit boundary layers. When $X_t$ takes only nonnegative values, Corollary  6.1 in \cite{Cetin2018} produces a boundary layer as $\xi\to \infty$. However, in our case, there can be two boundary layers as $\xi\to \pm \infty$. For $H\in \sC^1(\R)$, we illustrate that it is possible to have
\begin{align}\label{intro_FL}
\lim_{\xi\to\pm \infty}H'(\xi) \neq \lim_{t\downarrow 0}\lim_{\xi\to\pm \infty}h_x(t,\xi).
\end{align}
Unlike the smallness property characterization of $h^*$, such boundary layers are detected in our PDE characterization of  $h^*$  and produces explicit boundary values that can be used when truncating the state-space $\R$ to a finite grid (needed for finite difference methods).

%{ The paper is organized as follows:
%There are two key ingredients in our proof. First, we relate the strict local martingale property of the solution to \eqref{dX22intro} to growth properties of solutions to the corresponding (singular) Sturm-Lioville ODEs \eqref{SLODE} denoted by $\varphi_{\lambda\uparrow}$ and $\varphi_{\lambda\downarrow}$. Second, we construct suitable non-equivalent measures $(\P^{\varphi,x})_{x\in\R}$. 
%The third and last subsection contains the proof of Theorem \ref{thm_Main}.

{ 
\section{Assumptions and preliminaries}}
Let $\R_{\Delta}$ be the one-point compactification of $\R$ with $\Delta$ being the {\em point-at-infinity}. We consider the path space $\Omega$ of right continuous functions $\omega:[0,\infty)\to \R_{\Delta}$ which satisfy $\omega(t)=\Delta$ for all $t>s\ge0$ whenever $\omega(s)=\Delta$ and we note that $\sC(\R_+,\R)\subset \sC(\R_+,\R_{\Delta}) \subset \Omega$.  We let $X$ be the coordinate process on $\Omega$ and let $(\sF^0_t)_{t \geq 0}$ be its natural filtration $\sF^0_t := \sigma(X_u)_{u\in[0,t]}$. As in, e.g.,  Section 4 in  \cite{SharpeGTM}, we define the universal completion $\sF^*_t$ of $\sF^0_t$   by
\begin{align}
\sF^*_t := \cap_\mu \sigma (\sF^0_t, \mu\text{ null sets in }\sF_t^0),
\end{align}
where the intersection is taken over all finite measures $\mu$ on $\sF^0_t$. Finally, we define $\sF_t$ to be the right-continuous extension of $\sF^*_t$; that is, $\sF_t := \cap_{\epsilon>0} \sF^*_{t+\epsilon}$ { and we define  the $\sigma$-algebra} $\sF:=\bigvee_{t \geq 0} \sF_t$.  \ \\

{

\begin{assumption} \label{a:speed} The volatility function $\sigma:\R\to\R$ is Borel, $\sigma(x)^2 >0$ for all $x\in\R$, and for all $x\in \R$ there exists $\epsilon >0$ such that $\int_{x-\epsilon}^{x+\epsilon} \frac1{\sigma(y)^2} dy <\infty$.
$\endproof$
\end{assumption}
Assumption \ref{a:speed} is necessary and sufficient for  the existence of a unique weak solution $(\P^x)_{x\in\R}$ of \eqref{dX22intro}. This  result is due to Engelbert and Schmidt \cite{ES91} and a more recent account can be found in, e.g., Theorem 5.5.7 in \cite{KS88}.  Corollary 4.23 in \cite{ES91} also establishes that  $X$ is a strong Markov process under Assumption \ref{a:speed}.  Moreover, under Assumption \ref{a:speed}, }there are no absorbing states and therefore $X$ in \eqref{dX22intro} is recurrent under $(\P^x)_{x \in \R}$ in the sense that $\P^x(T_y<\infty)=1$ for all $(x,y)\in \R^2$ where 
\begin{align}\label{Ty}
T_y:=\inf\{t>0:X_t=y\},
\end{align}
{ see, e.g., Proposition 5.5.22(a) in Karatzas and Shreve (1988).  In particular, $X$'s  lifetime  $\zeta :=\inf\{t>0:X_{t} =\Delta\}$ satisfies $\P^x(\zeta<\infty)=0$ for all $x \in \R$ (see, e.g., Problem 5.5.3 in Karatzas and Shreve (1988)). Therefore, $X$ is a regular diffusion on $\R$ with speed measure 
\begin{align}\label{speed}
	m(d\xi):=  \frac{2}{\sigma(\xi)^2}d\xi,\quad \xi\in\R.
\end{align}

\subsection{Measure change on the path space}

}

Observe that the path space $\Omega$ is  projective (see Definitions 23.10 and 62.4 as well as the following paragraph in Sharpe \cite{SharpeGTM}). Thus, given a supermartingale multiplicative functional $Y=(Y_t)_{t\ge0}$ as defined in Eqs. (54.1) and (54.7) in \cite{SharpeGTM}, Corollary 62.21 in \cite{SharpeGTM} establishes the existence of Markov kernels $(\P^{Y,x})_{x\in\R_\Delta}$  $(\Omega, \sF)$ such that for any stopping time $T$, we have
\begin{align*}
\E^{Y,x}[F1_{T<\zeta}]=\E^x[F Y_T 1_{T<\zeta}], \quad F \in \mathbf{b}\sF_T,\quad x\in\R.
\end{align*}
{ That is,  $Y_T$ acts as a Radon-Nikodym derivative for changing the measure from $\P^x$ to $\P^{Y,x}$ on $\sF_T$ and $[T<\zeta]$.} 
We emphasize that while $\P^x(\zeta<\infty)=0$ for all $x \in \R$, the process $Y$ determines whether or not $\P^{Y,x}(\zeta=\infty)<1$ for all $x\in \R$.  

{ We shall change measure based on  $\lambda$-harmonic functions where a continuous function $u:\R\to (0,\infty)$ is called a  $\lambda$-harmonic function, $\lambda>0$, if the process 
$e^{-\lambda t}u(X_t)$, $t\ge0$, is a local martingale (see, e.g., Remark 2.3(iii) in Salminen and Ta \cite{ST2015}).  In that case, the normalized  process $Y_t:= e^{-\lambda t}\frac{u(X_t)}{u(X_0)}$, $t\ge0$, is a supermartingale multiplicative functional. Proposition \ref{p:harmonic} in the Appendix ensures that $u$ is both convex and $\lambda$-excessive.\footnote{The local martingale property only implies that $u$ is superaveraging in the sense that $u(x) \ge \E^x[u(X_t)]$ for $t\ge0$. However, as we show in Proposition \ref{p:harmonic}(ii) in Appendix \ref{appA}, $\lim_{t\downarrow  0}\E^x[u(X_t)]=u(x)$ holds automatically in our setting.} Paragraph II.30 in \cite{BorSal} ensures that any $\lambda$-harmonic function can be written as a linear combination of the \emph{minimal}  $\lambda$-harmonic functions  $\plu$ and $\pld$, where $\plu$ is strictly increasing and $\pld$ is strictly decreasing, strictly positive, and  satisfy
\begin{align}\label{SLODE0}
		\lambda \int_{[0,x)} u(\xi)m(d\xi)
		=u^-(x)-u^-(0),\quad x \in\R,
\end{align}
where $u^-$ is the left derivative and $m$ is from \eqref{speed}. Equivalently, $\plu$ and $\pld$ are uniquely given (up to multiplicative constants) by the Laplace transform properties:
\begin{align}\label{lambdaexx}
	\begin{split}
		\varphi_{\lambda \uparrow}(x) &= \varphi_{\lambda \uparrow}(y)\E^x[e^{-
			\lambda T_y}], \\
		\varphi_{\lambda \downarrow}(y) &= \varphi_{\lambda \downarrow}(x)\E^y[e^{-
			\lambda T_x}], 
	\end{split}
\end{align}
for $x,y\in \R$ with $x<y$, see, e.g., p.128 in It\^o and McKean \cite{IM1974} and Paragraph II.10 in Borodin and Salminen \cite{BorSal}. %From \eqref{SLODE0}, we see that $u$ is continuously differentiable since $m$ is absolutely continuous. Consistent with the terminology from the ODE literature, we shall call $\plu$ and $\pld$ the fundamental solutions of \eqref{SLODE0} (see Paragraph II.10 in \cite{BorSal} for further details). 

The change of measure via a $\lambda$-harmonic function and the corresponding characteristics of the resulting diffusion are given in} the next lemma, which we use repeatedly  in the sequel.

\begin{lemma}	\label{l:COM} Suppose Assumption \ref{a:speed} holds and let  $u:\R\to (0,\infty)$ be a $\lambda$-harmonic function for a constant $\lambda > 0$. Then:
	\begin{enumerate}
		\item[(i)] There exist  Markov kernels $(\P^{u,x})_{x\in\R_\Delta}$ such that for any stopping time $T$
	\begin{align} \label{e:COM}
	\E^{u,x}[F1_{T<\zeta}]=\frac1{u(x)}\E^x\big[F e^{-\lambda T}u(X_T) 1_{T<\zeta}\big], \quad F \in \mathbf{b}\sF_T,\quad x\in\R.
	\end{align}

	\item[(ii)]  
{ $u$ is strictly convex, $\lambda$-excessive, and satisfies \eqref{SLODE0}. If --- additionally --- $\sigma$ is continuous, then $u \in \mathcal{C}^2$ and satisfies the {\em Sturm-Liouville ODE}
\begin{align}\label{SLODE}
\lambda u(x)=	\frac{1}{2}\sigma^2(x) u''(x),\quad x\in \R.
\end{align}
\item[(iii)]   %{By denoting $u$'s left derivative by $u'$, the process
	%	\begin{align}\label{dtildeB}
%		\tilde{B}_t := B_t - \int_0^t \frac{u'(X_s)}{u(X_s)} \sigma(X_s)ds,\quad t\in[0,\zeta), \quad B_0^{u,x}:=0,
%		\end{align}
%		is a Brownian motion under $ \P^{u,x}$, $x\in \R$, and }
		$(\P^{u,x})_{x\in \R_\Delta}$ is the law of a diffusion with values in $\R_{\Delta}$, $\P^{u,x}(T_y <\infty) >0$ for all $x,y\in \R$,} a null killing measure, and scale function and speed measure given by
		\begin{align} \label{e:COMscale}
		s_u(z)= \int_0^z \frac{1}{u^2(\xi)}d\xi,\quad z\in \R,
		 \quad m_u(d\xi)=u^2(\xi)m(d\xi), \quad \xi \in \R.
		\end{align}
	 		\item[(iv)]  The mapping $(t,x)\mapsto \P^{u,x}(\zeta>t)$ is jointly continuous on $[0,\infty)\times \R$.
	\end{enumerate}
\end{lemma}
\proof
(i) This is a direct consequence of Corollary 62.21 in \cite{SharpeGTM}. 

 {
(ii) The first two statements follow from Proposition \ref{p:harmonic},  Paragraph II.30 in \cite{BorSal}, and that $\pld$ and $\plu$ satisfy \eqref{SLODE0}. The remaining assertion follows  from the continuity of $\sigma$ and \eqref{SLODE0}.

(iii) The formulas in \eqref{e:COMscale} can be found in Paragraph II.31 in \cite{BorSal}, Theorem 8.3 in Langer and Schenk \cite{LS90}, and  Theorem 6.2 in Evans and Hening (2019).} The killing measure under $\P^{u,x}$  is null because $u$ is a $\lambda$-harmonic function.

(iv)  Define the process $\tilde{X}_t:=s_u(X_t)$ for $t\ge0$. Then, $\tilde{X}$ is a  local martingale under $\P^{u,x}$ with volatility coefficient $s'_u\big(s_u^{-1}(\tilde{X}_t)\big)\sigma\big(s_u^{-1}(\tilde{X}_t)\big)$.  Proposition 4.3 in Karatzas and Ruf  \cite{KR16} gives continuity of $(t,x)\mapsto \P^{u,x}(\tilde{\zeta}>t)$ where $\tilde{\zeta} := \inf \big\{t>0 : \tilde{X}_t \in \{s_u(-\infty),s_u(\infty)\}\big\}$. Then, (iii) follows because $\tilde{\zeta} = \inf \{t>0 : X_t =\Delta \} =\zeta $.
$\endproof$

{
\subsection{Asymptotics of $\pld$ and $\plu$ }
}
The next result describes the asymptotic behavior of the minimal solutions $\pld$ and $\plu$ and is a consequence of  Theorem 2.2 in \cite{UZ17}.

\begin{theorem}[Urusov and Zervos \cite{UZ17}]\label{lem_general} { Under Assumption \ref{a:speed}, }the following are equivalent:

\begin{enumerate}
\item[(i)]  $\int_0^\infty \xi m(d\xi) <\infty$ (resp. $\int_{-\infty}^0 |\xi| m(d\xi) <\infty$).
\item[(ii)] $\varphi_{\lambda \uparrow}(\xi)$ (resp. $\varphi_{\lambda \downarrow}(\xi)$) has linear growth at $\xi=\infty$ (resp.  $\xi=-\infty)$. 
\item[(iii)]The process $\big(e^{-\lambda t} \varphi_{\lambda \uparrow}(X_t)\big)_{t\ge0}$ (resp. $\big(e^{-\lambda t} \varphi_{\lambda \downarrow}(X_t)\big)_{t\ge0}$) is a strict  local martingale under $\P^x$ for all $x\in\R$. 
\end{enumerate}
\end{theorem}
\proof  First, the boundary point $\infty$ is inaccessible because
\begin{align}
\int_0^{\infty} \int_0^y m(d\xi) dy=\infty.
\end{align}
Furthermore, the  point $\infty$ is a natural or entrance boundary depending on whether
\begin{align}
\begin{split}
\int_0^{\infty} m((y,\infty)) dy&=\int_0^{\infty} \int_0^\xi dy m(d\xi) \\
&=\int_0^{\infty} \xi m(d\xi) 
\end{split}
\end{align}
is infinite or not. Thus, the claimed equivalences follow from  Theorem 2.2 in \cite{UZ17}.

$\endproof$

In the setting of the  logarithm of the two dimensional Bessel process, the next example gives { $\varphi_{\lambda \uparrow}$ and  $\varphi_{\lambda \downarrow}$ explicitly}.

\begin{example} \label{bes2D}For $\sigma(\xi) := e^{-\xi}$, $\xi\in\R$, the dynamics \eqref{dX22intro} become those of the logarithm of the two dimensional Bessel process that solves the SDE 
\begin{align}\label{dX2}
dX_t = e^{-X_t}dB_t,\quad t\ge 0, \quad X_0=x \in \R.
%d\log(R_t) = \frac1{R_t} dR_t - \frac12 \frac1{R_t^2} dt=\frac1{R_t}dB_t.
\end{align}
Eq. \eqref{dX2} has a unique strong solution for a given Brownian motion $B = (B_t)_{t\ge0}$.
 For a constant $\lambda>0$, the corresponding minimal solutions of \eqref{SLODE0} are
\begin{align}\label{2dphi}
\begin{split}
&\varphi_{\lambda \uparrow}(\xi):= I_0(e^\xi\sqrt{2\lambda}),\quad \xi\in\R,\quad \lim_{\xi\uparrow\infty}\frac{\varphi_{\lambda \uparrow}(\xi)}{\xi} = \infty,\\% \sim \frac{e^{e^\xi\sqrt{2\lambda}}}{\sqrt{2\pi e^\xi\sqrt{2\lambda}}} \text{ for } \xi\to\infty,\\
&\varphi_{\lambda \downarrow}(\xi):= K_0(e^\xi\sqrt{2\lambda}),\quad \xi\in\R,\quad \lim_{\xi\downarrow-\infty}\frac{\varphi_{\lambda \downarrow}(\xi)}{\xi} = -1,%\quad \varphi_{\lambda \downarrow}(\xi) \sim %\sqrt{\frac{\pi}{2e^x\sqrt{2\lambda}}} e^{-e^x\sqrt{2\lambda}} -\xi-\log(\sqrt{2\lambda}) \text{ for } \xi\to-\infty,
\end{split}
\end{align}
see, e.g., Jeanblanc, Yor, and Chesney (\cite{JYC2009}, p.279). In \eqref{2dphi}, the functions $I_0$ and $K_0$ are modified Bessel functions. The superlinear growth limit in \eqref{2dphi} (as $\xi \uparrow \infty)$ {and Theorem \ref{lem_general}} ensure that the local martingale $\big(e^{-\lambda t}\varphi_{\lambda \uparrow}(X_t)\big)_{t\ge0}$ is a martingale.
The linear growth limit in \eqref{2dphi} (as $\xi \downarrow -\infty)$ {and Theorem \ref{lem_general}} ensure that the local martingale  $\big(e^{-\lambda t}\varphi_{\lambda \downarrow}(X_t)\big)_{t\ge0} $ is a strict local martingale. $\endproof$
\end{example}

{
\subsection{Asymptotics of the expectation of $\pld$ and $\plu$ }
%Lemma \ref{cetin2018} below, in particular   the limits \eqref{finfty2} and \eqref{finfty3}, can be seen as an extension of Corollary 6.1 in \c{C}et\.in \cite{Cetin2018}  to the minimal $\lambda$-harmonic functions which 
In the next section, we need the following result.}   

\begin{lemma}\label{cetin2018} Suppose Assumption \ref{a:speed} holds.

\begin{enumerate}
\item[(i)]  If $\int_0^\infty \xi m(d\xi) <\infty$, the strict local martingale 
\begin{align}\label{Yup}
Y_t:= e^{-\lambda t} \varphi_{\lambda\uparrow}(X_t),\quad t\ge0, \quad \lambda>0,
\end{align}
has the asymptotic expectation
\begin{align}\label{finfty2}
\lim_{n\uparrow\infty}\frac{\E^{n}[Y_{s_n}]}{\varphi_{\lambda\uparrow}(n)}=\lim_{n\uparrow\infty}\frac{\E^{n}[Y_{s_n}]}{n}=0,\quad \infty>s_n\ge s_{n+1},\quad s_\infty:= \lim_{n\uparrow\infty} s_n >0.
\end{align}
\item[(ii)] If $\int_{-\infty}^0 |\xi| m(d\xi) <\infty$, the strict local martingale 
\begin{align}\label{Ydown}
Y_t:= e^{-\lambda t} \varphi_{\lambda\downarrow}(X_t),\quad t\ge0, \quad \lambda>0,
\end{align}
has the asymptotic expectation
\begin{align}\label{finfty3}
\lim_{n\uparrow\infty}\frac{\E^{n}[Y_{s_n}]}{\varphi_{\lambda\downarrow}(-n)}=\lim_{n\uparrow\infty}\frac{\E^{n}[Y_{s_n}]}{n}=0,\quad \infty>s_n\ge s_{n+1},\quad s_\infty:= \lim_{n\uparrow\infty} s_n >0.
\end{align}
\end{enumerate}

\end{lemma}

 \proof
\noindent (i): {\bf Step 1/3:} From Theorem \ref{lem_general}, we know that \eqref{Yup} is a strict local martingale and $\plu(\xi)$ is of linear growth as $\xi\uparrow\infty$. It follows from Lemma \ref{l:COM} that there exist Markov kernels $(\P^{\varphi,x})_{x\in \R_{\Delta}}$ such that \eqref{e:COM} holds with $u:=\plu$. Moreover, 
\begin{align}\label{slambdaup}
s_{\lambda\uparrow}(z):=\int_0^z \frac{1}{\plu^2(\xi)}d\xi,\quad z\in \R,\quad  m_{\lambda\uparrow}(d\xi) :=2\frac{\varphi_{\lambda \uparrow}(\xi)^2}{\sigma(\xi)^2}d\xi,\quad \xi\in\R,
\end{align}
are a scale function and a speed measure for $X$ under $(\P^{\varphi,x})_{x\in \R_{\Delta}}$. Since $s_{\lambda\uparrow}(-\infty)=-\infty$ and $s_{\lambda\uparrow}(\infty)<\infty$, the limit $X_{\infty}$ exists and $\P^{\varphi,x}(X_\infty =\infty)=1$ for all $x\in \R$ {(see, e.g., Proposition 5.5.22(c) in \cite{KS88})}.

\noindent {\bf Step 2/3:} Define the stopping times 
\begin{align}\label{taun}
\begin{split}
T_n &:= \inf\{t> 0 : X_t \ge n\},\quad n\in\N,
\end{split}
\end{align}
and observe that $T_n$ increases to $X$'s lifetime $\zeta$ under $\P^{\varphi,x}$ as $n\uparrow \infty$. Moreover, $\P^x(T_n<\infty)=1$ for all $n\in\N$ due to $X$'s recurrence. Then, for $t\in [0,\infty)$, Lemma  \ref{l:COM} yields 
\begin{align}\label{dom1}
\begin{split}
\P^{\varphi,x}(t<T_n) &=\tfrac{1}{Y_0}\E^{x}[Y_{T_n}1_{t<T_n}] \\
  &=\tfrac{1}{\varphi_{\lambda\uparrow}(x)}\E^{x}[Y_t1_{t<T_n} ],
\end{split}
\end{align}
where the last equality is due to the fact that $(Y_{t\land T_n})_{t\ge0}$ is a uniformly integrable martingale for $n\in\N$. Because $\P^{x}(\lim_{n \uparrow \infty} T_n=\infty)=1$, the dominated convergence theorem  {applied to \eqref{dom1} }gives us 
\begin{align}\label{defw}
\begin{split}
w(t,x)=\tfrac{1}{\varphi_{\lambda\uparrow}(x)}\E^{x}[Y_t], 
\end{split}
\end{align}
where we have defined the function 
\begin{align}\label{defww}
\begin{split}
w(t,x):= \P^{\varphi,x}(t<\zeta),\quad t\ge0,\quad x\in\R.
\end{split}
\end{align}
{ Lemma \ref{l:COM}(iv) ensures that the function $w$ in \eqref {defww} is  jointly continuous.}

\noindent {\bf Step 3/3:} For $n\in\N$, let $T_n$ be defined in \eqref{taun} and let $t\in[0,\infty)$. Then,  we have 
\begin{align}\label{shiftA4}
\begin{split}
\P^{\varphi,x} (\zeta >t) &=  \P^{\varphi,x} (\zeta >t, T_n \le t)+\P^{\varphi,x} (\zeta >t, T_n >t)\\
&= \E^{\varphi,x}\Big[1_{T_n\le t}\P^{\varphi,X_{T_n}} (\zeta >t-u)|_{u=T_n}\Big]+\P^{\varphi,x}  (T_n>t)\\
&=\E^{\varphi,x}\Big[1_{T_n\le t}w\big(t-T_n,n\big)\Big]+\P^{\varphi,x}  (T_n>t),
\end{split}
\end{align}
where the second line follows from the strong Markov property of $X$ under $\P^{\varphi,x}$ in view of Lemma \ref{l:COM} (see, e.g.,  Exercise 6.12 in \cite{SharpeGTM}).

Let $(s_n)_{n\in\N} \subset (0,\infty)$ be a non-increasing sequence with a positive limit $s_\infty \in (0,\infty)$. By replacing $t$ with $s_n$ in \eqref{shiftA4}, we can use dominated convergence when passing $n\uparrow \infty$ in \eqref{shiftA4} to see
\begin{align}\label{shiftA}
\begin{split}
0&=\lim_{n\uparrow \infty} \E^{\varphi,x}\Big[1_{T_n\le s_n}w\big(s_n-T_n,n\big)\Big]\\
&=\E^{\varphi,x}\Big[1_{\zeta\le s_\infty}\lim_{n\uparrow \infty}w\big(s_n-T_n,n\big)\Big].
\end{split}
\end{align}
Therefore, because $w\ge0$, we have
\begin{align}\label{shiftAAA}
\begin{split}
1_{\zeta\le s_\infty}\lim_{n\uparrow \infty}w\big(s_n-T_n,n\big)=0,\quad \P^{\varphi,x} \text{-a.s.}
\end{split}
\end{align}

By using \eqref{defw} and the strict local martingale property of \eqref{Yup} we see that $w(t,x) <1$ for $t\in(0,\infty)$ and $x\in\R$. Therefore, because $s_\infty\in(0,\infty)$, we have
\begin{align}\label{shiftA1}
\begin{split}
\P^{\varphi,x}(\zeta\le s_\infty )&=1-w(s_\infty,x) %\\&=1-\tfrac{1}{\varphi_{\lambda\uparrow}(x)}\E^{x}[Y_{s_\infty}]\\&
>0.
\end{split}
\end{align}
Consequently, the set  $(\zeta\le s_\infty)$ {in \eqref{shiftAAA}} is not $\P^{\varphi,x}$-null. Because $s_n$ is non-increasing with limit $s_\infty\in(0,\infty)$ and $T_n(\omega)$ is non-decreasing with limit $\zeta(\omega)$  for $\omega\in (\zeta\le s_\infty)$, we have
$$
T_n(\omega) \le \zeta (\omega)\le s_\infty \le s_n,\quad \omega\in (\zeta\le s_\infty).
$$
Thus, 
\begin{align}\label{e:limwn}
1_{\zeta\le s_\infty}\lim_{n\uparrow \infty}w\big(s_n-T_n,n\big)\geq 1_{\zeta\le s_\infty}\lim_{n\uparrow \infty}w\big(s_n,n\big),
\end{align}
where the inequality uses that $t\to w(t,x)$ is non-increasing for all $x\in\R$. Combining \eqref{shiftAAA} and \eqref{e:limwn} yields $\lim_{n\uparrow \infty}w\big(s_n,n\big)=0$ because  the set  $(\zeta\le s_\infty)$ is not $\P^{\varphi,x}$-null. Finally, using \eqref{defw}, we arrive at
$$
0=\lim_{n\uparrow \infty}w(s_n,n)=\lim_{n\uparrow \infty} \tfrac{1}{\varphi_{\lambda\uparrow}(n)}\E^{n}[Y_{s_n}].
$$

\noindent (ii): This proof is similar to the proof of (i) and is omitted.

$\endproof$

{	In the setting of Section 6 of \c{C}et\.in \cite{Cetin2018},  $\varphi_0(x):=x$, $x\ge0$, is a minimal harmonic function (unique up to a multiplicative constant) because $X_t$ is a nonnegative local martingale. A special case of Corollary 6.1 in \cite{Cetin2018} shows that $\lim_{n\rar \infty}\frac{\E^n[\varphi_0(X_t)]}{n}=0$, which is the analogue of the limits  \eqref{finfty2} and \eqref{finfty3} associated with the minimal $\lambda$-harmonic functions $\varphi_{\lambda \uparrow}$ and $\varphi_{\lambda \downarrow}$ for $\lambda>0$ and $s_n:=t$ for $n\in\N$.}

The following example shows that the limits  \eqref{finfty2} and \eqref{finfty3} do not hold for arbitrary { positive} strict local martingales.

\begin{example}\label{exxx3Dbesselll} {
Let $(Y_t)_{t\ge0}$ be the inverse three dimensional Bessel process with dynamics 
\begin{align}\label{dX3d}
dY_t := -Y_t^2dB_t,\quad t\in( 0,\infty), \quad Y_0=y>0.
\end{align}
Eq. \eqref{dX3d} has a unique strong solution $Y^y = (Y^y_t)_{t\ge 0}$ for any initial value $y\in(0,\infty)$ and for a given Brownian motion $B=(B_t)_{t\ge0}$. This is the classical example due to Johnson and Helms \cite{JH63} of a strict local martingale. }Then, the strict local martingale  $\tilde{Y}_t :=\tilde{y} Y^1_t$ for $t\ge0$ and $\tilde{y}>0$ satisfies
\begin{align}\label{ex3Dbesselll}
\lim_{\tilde{y}\uparrow\infty}\frac{\E^{\P}[\tilde{Y}_{t}]}{\tilde{y}} =\E^{\P}[Y^1_t] \in(0,1),\quad t\in (0,\infty).
\end{align}
We note that the dependence on the initial value $\tilde{y}>0$ in the dynamics $d\tilde{Y}_t = - \frac1{\tilde{y}} \tilde{Y}_t^2dB_t$ implies that Corollary 6.1 in \c{C}et\.in (2018) cannot be applied. 
$\endproof$
\end{example}

\section{Main results and examples}

{ In the first subsection, our first result relies only on a unique weak solution of the SDE \eqref{dX22intro}. The second result uses a stronger local $\frac12$-H\"older continuity and the resulting unique strong solution of \eqref{dX22intro}.  The second subsection gives two examples.

}

\subsection{Main results}
{ Our first main result gives uniqueness of weak and classical solutions and existence of a weak solution of \eqref{CauchyI1}. The class of functions in which we establish uniqueness is constructed based on Theorem 1 in Kotani \cite{Kotani2006} and Lemma \ref{cetin2018} above. The former result} ensures that when the solution of \eqref{dX22intro} is a strict local martingale, at least one of (i) $\int_0^\infty \xi m(d\xi) <\infty$ and (ii) $\int_{-\infty}^0 |\xi| m(d\xi) <\infty$ holds (see also Theorem 1.6 in Delbaen and Shirakawa \cite{DS2002} for the case when the solution of \eqref{dX22intro} is  nonnegative).  In terms of { $\varphi_{\lambda \uparrow}$ and $\varphi_{\lambda \downarrow}$ from \eqref{lambdaexx}}, we can define the nonnegative convex function
\begin{align}\label{PHI}
\Phi_\lambda:= \varphi_{\lambda\downarrow}+\varphi_{\lambda\uparrow},
\end{align}
which is uniformly bounded away from zero. { Because both $\varphi_{\lambda\downarrow}$ and $ \varphi_{\lambda\uparrow}$ are $\lambda$-harmonic functions, so is $\Phi_\lambda$. 

We write $f\in \mathcal{H}_{\lambda\uparrow}$ if $f:[0,\infty)\times \R\to \R$ is continuous and satisfies $f(0,x) = H(x), \; |f(t,x)|\le Ke^{\lambda t}\Phi_\lambda(x)$ for some constant $K\ge0$ ($K$ can vary with $f$), and $f$ satisfies the boundary condition
\begin{align}\label{UC2}
\lim_{n\uparrow \infty} \frac{f(s_n,n)}{n}=0,\quad \text{whenever}\quad \infty >  s_n \ge s_{n+1} \text{ and }s_\infty:=\lim_{n\uparrow \infty} s_n >0.
\end{align}
The set $\mathcal{H}_{\lambda\downarrow}$ is defined similarly, except we replace \eqref{UC2} with the boundary condition
\begin{align}\label{UC3}
\lim_{n\uparrow \infty} \frac{f(s_n,-n)}{n}=0,\quad \text{whenever}\quad\infty > s_n \ge s_{n+1} \text{ and }s_\infty:=\lim_{n\uparrow \infty} s_n >0.
\end{align}

}

\begin{theorem}\label{thm_Main} Suppose Assumption \ref{a:speed} holds. Then, for continuous data $H:\R\to\R$ such that 
\begin{align}\label{Hbound1}
\sup_{\xi, |\xi|>1}\frac{|H(\xi)|}{\Phi_\lambda(\xi)}<\infty,
%|H(\xi)| \le d_0 + d\Phi_\lambda(\xi),\quad \xi \in\R,
\end{align}
{ and recall  $h^*$ from \eqref{def_h}. Then, we have:}
\begin{itemize}

\item[(i)]  If $\int_0^\infty \xi m(d\xi) <\infty$ and $\int_{-\infty}^0 |\xi| m(d\xi)=\infty$, { $h^*\in \mathcal{H}_{\lambda\uparrow}$, and  there is at most one classical solution $h\in \sC^{1,2}$ of \eqref{CauchyI1} in $ \mathcal{H}_{\lambda\uparrow}$. Moreover,  $h^*$ is the unique  weak solution   of \eqref{CauchyI1} in  $\mathcal{H}_{\lambda\uparrow}$.} 

\item[(ii)]  If $\int_0 ^\infty\xi m(d\xi)=\infty$ and $\int_{-\infty} ^0|\xi| m(d\xi) <\infty$,  { $h^*\in \mathcal{H}_{\lambda\downarrow}$, and there is at most one classical solution $h\in \sC^{1,2}$ of \eqref{CauchyI1} in $ \mathcal{H}_{\lambda\uparrow}$.  Moreover,  $h^*$ is the unique  weak solution   of \eqref{CauchyI1} in  $\mathcal{H}_{\lambda\downarrow}$. } 

\item[(iii)]  If $\int_0^\infty \xi m(d\xi) <\infty$ and  $\int_{-\infty} ^0|\xi| m(d\xi) <\infty$,  { $h^*\in \mathcal{H}_{\lambda\uparrow}\cap \mathcal{H}_{\lambda\downarrow}$, and  there is at most one classical solution $h\in \sC^{1,2}$ of \eqref{CauchyI1} in $ \mathcal{H}_{\lambda\uparrow} \cap \mathcal{H}_{\lambda\uparrow}$.  Moreover,  $h^*$ is the unique  weak solution   of \eqref{CauchyI1}  in  $\mathcal{H}_{\lambda\uparrow}\cap \mathcal{H}_{\lambda\downarrow}$}. 
\end{itemize}
\end{theorem}

\noindent \emph{Proof of (i):} {  {\bf Step 1/2:} First, because $\Phi_{\lambda}\in \sC$, we see that $h^*$ is continuous on $[0,\infty)\times \R$  if and only if $\E^{\Phi,x}[\frac{H(X_t)}{\Phi_{\lambda}(X_t)}1_{t<\zeta}]$ is continuous.  We define $\bar{h}(x):=\frac{H(x)}{\Phi_{\lambda}(x)}$ for $x\in \R$. We let $x_n\rightarrow x \in \R$ ,$t_n \rightarrow t \geq 0$, and define the sequence of stopping times  $T_n:=\inf\{t\geq 0: X_t=x_n\}$. Blumenthal's 0-1 law gives $\P^{\Phi,x}(\lim_{n\rightarrow\infty}T_n=0)=1$.  Because $\bar{h}\in \sC_b$, we have 
	\begin{align*}
		\E^{\Phi,x}[\bar{h}(X_{t_n+T_n})1_{t_n+T_n<\zeta}1_{T_n<\zeta}]&=\E^{\Phi,x}[\E^{\Phi,x_n}[\bar{h}(X_{t_n})1_{t_n<\zeta}]1_{T_n<\zeta}]\\
		&=\E^{\Phi,x_n}[\bar{h}(X_{t_n})1_{t_n<\zeta}]\P^{\Phi,x}(T_n<\zeta),
	\end{align*} 
	where the first equality is due to the strong Markov property. Since $\bar{h}$ is bounded, $T_n\to0$, $\P^{\Phi,x}$-a.s., and $\P^{\Phi,x}(\zeta>0)=1$, the dominated convergence theorem gives
	\[
	\E^{\Phi,x}[\bar{h}(X_{t})1_{t<\zeta}]=\lim_{n\rightarrow \infty}\E^{\Phi,x_n}[\bar{h}(X_{t_n})1_{t_n<\zeta}]\P^{\Phi,x}(T_n<\zeta)=\lim_{n\rightarrow\infty}\E^{\Phi,x_n}[\bar{h}(X_{t_n})1_{t<\zeta}].
	\]
This completes the proof of joint continuity.   Moreover, since $h^*(t,x)=\E^x[H(X_t)]$,  the process $\big(h^*(T-t, X_t)\big)_{t\in [0,T]}$ is a $\P^x$-martingale for any $T>0$, which in turn implies that $h^*$ is a weak solution by Sawyer \cite{Sawyer} (see Theorem \ref{t:weak} in Appendix \ref{s:weak} for details). The remaining properties needed for $h^* \in \mathcal{H}_{\lambda\uparrow}$ follow from \eqref{Hbound1} and Lemma \ref{cetin2018}. 

 {\bf Step 2/2:} Let $h \in \mathcal{H}_{\lambda\uparrow}$ be a  classical or weak solution of  \eqref{CauchyI1}. 
}
 
 For a fixed constant $T\in[0,\infty)$, we establish the uniqueness claims by proving the following representation
\begin{align}\label{htilde}
\begin{split}
\E^{\Phi,x}\left[\frac{H(X_T)}{e^{-\lambda T}\Phi_\lambda(X_T)}1_{T<\zeta}\right] 
  &=\tilde{h}(T,x),\quad x\in\R,
\end{split}
\end{align}
where $\tilde{h}$ is defined as
\begin{align}\label{tildeh}
\tilde{h}(t,x):=\frac{h(t,x)}{\Phi_\lambda(x)},\quad t\ge0,\quad x\in\R,
\end{align} 
and the Markov kernels $(\P^{\Phi,x})_{x\in \R_{\Delta}}$ are from Lemma \ref{l:COM} with $u:=\Phi_\lambda$ where $\Phi_\lambda$ is defined in \eqref{PHI}. { By assumption, the function $\tilde h$ in \eqref{tildeh} satisfies the uniform bound
\begin{align}\label{htboundA}
|\tilde{h}(t,x)| \le Ke^{\lambda T},\quad t\in[0,T],\quad x\in \R.
\end{align}
}Similarly to \eqref{slambdaup}, the scale function associated with the diffusion $X$ under $\P^{\Phi,x}$ is given by
\begin{align}\label{slambdaup2}
s_\Phi(z):=\int_0^z \frac{1}{\Phi_\lambda^2(\xi)}d\xi,\quad z\in \R.
\end{align}
Because $\lim_{z\downarrow-\infty}s_\Phi(z)>-\infty$ and $\lim_{z\uparrow\infty}s_\Phi(z)<\infty$, the limit of $X$ satisfies $X_{\infty}\in \{-\infty,\infty\}$, $\P^{\Phi,x}$-a.s., for $x\in \R$ { (see, e.g., Proposition 5.5.22(d) in \cite{KS88})}. We also introduce the stopping times
\begin{align}\label{num}
\begin{split}
\nu_n &:= \inf\{t> 0 : |X_t| \ge n\},\quad n\in\N,
\end{split}
\end{align}
and observe that $\lim_{n\uparrow\infty}\nu_n = \zeta,$ $\P^{\Phi,x}$-a.s., for all $x\in \R$ where we recall that $\zeta$ denotes $X$'s lifetime.

{ To see that \eqref{htilde} holds, we note that on $\sF_{t\land \nu_n}$, $\frac{d\P^{\Phi,x}}{d\P} = e^{-\lambda t \land \nu_n}\frac{ \Phi_\lambda(X_{ t\land \nu_n})}{\Phi_\lambda(x)}$ for $n\in \N$ and $t\in [0,\infty)$. Because 
\begin{align}
\tilde{h}(T-t\land \nu_n,X_{t\land \nu_n}) \frac{ \Phi_\lambda(X_{ t\land \nu_n})}{\Phi_\lambda(x)} = h(T-t\land \nu_n,X_{t\land \nu_n})\frac{1}{\Phi_\lambda(x)}
\end{align}
%we use It\^o's lemma to produce the local martingale dynamics under  $\P^{\Phi,x}$
%\begin{align}
%\begin{split}
%de^{\lambda t}\tilde{h}(T-t,X_t) &=e^{\lambda t}\Big((\lambda \tilde h-\tilde{h}_t)dt +\tilde h_x \sigma dB_t+ \frac12 \tilde h_{xx} \sigma^2dt \Big) \\
%&=e^{\lambda t}\Big((\lambda \tilde h-\tilde{h}_t)dt +\tilde h_x \sigma \Big(d\tilde{B}_t + \sigma\frac{\Phi'}{\Phi}dt\Big)+ \frac12 \tilde h_{xx} \sigma^2dt \Big) \\
%&=e^{\lambda t}\tilde h_x \sigma d\tilde{B}_t.
%\end{split}
%\end{align}
%The last equality follows from  the definition of $\tilde h$ in \eqref{tildeh}, the PDE in \eqref{CauchyI1}, the Sturm-Liouville ODE in \eqref{SLODE}, and the Brownian motion $\tilde{B}_t$ in \eqref{dtildeB}.   
and $h(T-t\land \nu_n,X_{t\land \nu_n})$ is a martingale under $\P^x$,  
the process $e^{\lambda t\land \nu_n}\tilde{h}(T-t\land \nu_n,X_{t\land \nu_n})$ is a martingale under $\P^{\Phi,x}$. This is because when $h$ is a classical solution of \eqref{CauchyI1}, It\^o's lemma gives the local martingale property under $\P^x$ and when $h$ is a weak solution of \eqref{CauchyI1}, the local martingale property under $\P^x$ follows from Sawyer \cite{Sawyer} (see also Lemma \ref{t:weak} in Appendix \ref{s:weak}).}
This gives us
\begin{align}\label{limit1}
\begin{split}
\tilde{h}(T,x) &= \E^{\Phi,x}\big[e^{\lambda T\land \nu_n}\tilde{h}(T-T\land \nu_n,X_{T\land \nu_n})\big]\\
&=  \E^{\Phi,x}\big[e^{\lambda T}\tilde{h}(0,X_T)1_{T<\nu_n}\big]+ \E^{\Phi,x}\big[e^{\lambda \nu_n}\tilde{h}(T-\nu_n,X_{\nu_n})1_{T\ge \nu_n}\big].
\end{split}
\end{align}
{The bound \eqref{htboundA} allows us to }use dominated convergence in \eqref{limit1} when passing  $n\uparrow \infty$ to see that
\begin{align}
\begin{split}
\tilde{h}(T,x)&= \E^{\Phi,x}\big[e^{\lambda T}\tilde{h}(0,X_T)1_{T<\zeta}\big]+\lim_{n\uparrow \infty} \E^{\Phi,x}\big[e^{\lambda \nu_n}\tilde{h}(T-\nu_n,X_{\nu_n})1_{T\ge \nu_n}\big]
\\
&= \E^{\Phi,x}\left[\frac{e^{\lambda T}H(X_T)}{\Phi_\lambda(X_T)}1_{T<\zeta}\right]+ \E^{\Phi,x}\big[\lim_{n\uparrow \infty}e^{\lambda \nu_n}\tilde{h}(T-\nu_n,X_{\nu_n})1_{T\ge \nu_n}\big],
\end{split}
\end{align}
where the second equality uses the initial condition $\tilde{h}(0,x)=\frac{H(x)}{\Phi_\lambda(x)}$ { from \eqref{tildeh}}. 
 Therefore, the representation in \eqref{htilde} follows as soon as we show 
\begin{align} \label{e:htildlim}
\lim_{n\uparrow \infty}\tilde{h}(T-\nu_n,X_{\nu_n})1_{T\ge \nu_n}=0,\quad \P^{\Phi,x}\text{-a.s.}
\end{align}

First, on the set $(T\ge \nu_n)$, { \eqref{num} gives } $X_{\nu_n} \in \{-n,n\}$. Therefore, 
\begin{align}\label{LimitA}
\begin{split}
\varphi_{\lambda \downarrow}(x) &= \lim_{n\uparrow \infty}\E^x\big[ e^{-\lambda T\land \nu_n} \varphi_{\lambda \downarrow}(X_{T\land \nu_n})(1_{T\ge \nu_n} +1_{T< \nu_n})\big]\\
&\ge \lim_{n\uparrow \infty}\Big(\varphi_{\lambda \downarrow}(-n)\E^x[ e^{-\lambda T\land \nu_n} 1_{T\ge \nu_n}1_{X_{\nu_n}=-n}]+ \E^x\big[ e^{-\lambda T\land \nu_n} \varphi_{\lambda \downarrow}(X_{T\land \nu_n})1_{T< \nu_n}\big]\Big)\\
&= \lim_{n\uparrow \infty}\Big(\frac{\varphi_{\lambda \downarrow}(-n)}{\Phi_\lambda(-n)}\E^x\big[ e^{-\lambda T\land \nu_n} \Phi_\lambda(X_{T\land \nu_n})1_{T\ge \nu_n}1_{X_{\nu_n}=-n}\big]+ \E^x\big[ e^{-\lambda T} \varphi_{\lambda \downarrow}(X_T)1_{T< \nu_n}\big]\Big)\\
&=  \Phi_\lambda(x) \lim_{n\uparrow \infty}\frac{\varphi_{\lambda \downarrow}(-n)}{\Phi_\lambda(-n)}\P^{\Phi,x}(T\ge \nu_n, X_{\nu_n}=-n )+ \E^x\big[ e^{-\lambda T} \varphi_{\lambda \downarrow}(X_T)\big]\\
&=  \Phi_\lambda(x)\lim_{n\uparrow \infty}\P^{\Phi,x}( T\ge \nu_n, X_{\nu_n}=-n )+\varphi_{\lambda \downarrow}(x).
\end{split}
\end{align}
{The second last equality uses dominated convergence. The last equality uses  the martingale property of $e^{-\lambda t} \varphi_{\lambda \downarrow}(X_t)$ from Theorem \ref{lem_general} and 
$$
 \lim_{n\uparrow \infty}\frac{\varphi_{\lambda \downarrow}(-n)}{\Phi_\lambda(-n)}=\lim_{n\uparrow \infty}\frac{\varphi_{\lambda \downarrow}(-n)}{\varphi_{\lambda \uparrow}(-n)+\varphi_{\lambda \downarrow}(-n)}=1
 $$ 
 because $\lim_{n\uparrow \infty}\varphi_{\lambda \uparrow}(-n) \in [0,\infty)$. The zero limit in \eqref{LimitA} and the bound \eqref{htboundA} give} 
\begin{align}\label{obs1}
\begin{split}
\lim_{n\uparrow \infty}\tilde{h}(T-\nu_n,-n)1_{T\ge \nu_n}1_{X_{\nu_n}=-n}=0, \quad \P^{\Phi,x}\text{-a.s.}
\end{split}
\end{align}

Second, because the set  $(\zeta =T)$ is $\P^{\Phi,x}$-null by Lemma \ref{l:COM}(iv), the sets $(T\ge \zeta )$ and $(T>\zeta )$ differ only by a $\P^{\Phi,x}$-null set. Then, we can use the boundary condition \eqref{UC2} and the linear growth of $\varphi_{\lambda\uparrow}(\xi)$ and $\Phi_{\lambda}(\xi)$ as $\xi \uparrow \infty$ to see
\begin{align}\label{obs2}
\begin{split}
\lim_{n\uparrow \infty}\tilde{h}(T-\nu_n,n)1_{T\ge \nu_n}1_{X_{\nu_n}=n}&=1_{T\ge \zeta}1_{X_{\zeta}=\infty}\lim_{n\uparrow \infty}\tilde{h}(T-\nu_n,n)\\
&=1_{T> \zeta}1_{X_{\zeta}=\infty}\lim_{n\uparrow \infty}\tilde{h}(T-\nu_n,n)\\
&=0,
\end{split}
\end{align}
$ \P^{\Phi,x}\text{-a.s.}
$
The two observations \eqref{obs1} and \eqref{obs2} establish \eqref{e:htildlim}.\ \\

\noindent (ii) and (iii): These are similar to (i) and are omitted. 

$\endproof$

We note that \eqref{Hbound1} covers continuous data $H:\R\to\R$ of at most linear growth, i.e.,
\begin{align}\label{Hlinear}
\sup_{\xi, |\xi|>1}\frac{|H(\xi)|}{|\xi|}<\infty.
%|H(\xi)| \le D'_0 + D'|\xi|
\end{align}
When $H$ satisfies \eqref{Hlinear}, Theorem 3.10(i) in \cite{HP2011} ensures that $h^*$ in \eqref{def_h} is also of at most linear growth. { However, our condition \eqref{Hbound1} is more general than \eqref{Hlinear} because Theorem \ref{lem_general}(ii) shows that } when $\int_{-\infty}^0 |\xi| m(d\xi) =\infty$,  \eqref{Hbound1}  allows for superlinearly growing data $H$ as $\xi\downarrow-\infty$ and when  $\int_0^\infty \xi m(d\xi) =\infty$ \eqref{Hbound1}   allows for superlinearly growing data $H$ as $\xi\uparrow \infty$.  For example,  the  logarithm of the two dimensional Bessel process in Example \ref{bes2D} allows for  superlinearly growing data as $\xi\uparrow \infty$. 

{Because $H(\xi):=\xi$ satisfies \eqref{Hlinear}, Assumption \ref{a:speed} ensures that $\E^x[|X_t|]<\infty$ for $x\in\R$ and $t\ge0$ (this also follows from Lemma 1 in \cite{Kotani2006}). However, }the following example shows that real-valued strict local martingales can fail to be integrable in general.
\begin{example}{ Let $(Y_t)_{t\ge0}$ denote the inverse three dimensional Bessel process with dynamics \eqref{dX3d} and initial value $y\in(0,\infty)$.} From, e.g., p.74 in Protter \cite{Pro}, the second moment satisfies  $\E^y[Y_t^2]<\infty$ while $\E^y[\langle Y\rangle_t]=\infty$ for $t\in(0,\infty)$.  Consequently, the real-valued  local martingale
\begin{align}
X_t:= Y_t^2 -\langle Y\rangle_t,\quad t\ge0,
\end{align}
is not integrable. In particular, $(X_t)_{t\ge0}$ is a strict local martingale too.
$\endproof$
\end{example}

{Under the following stronger local H\"older-continuity assumption on $\sigma$, our next result ensures existence of a classical solution to   \eqref{dX22intro}.

	\begin{assumption} \label{a:holder} The volatility function $\sigma:\R\to(0,\infty)$ is locally $\frac12$-H\"older continuous.
		$\endproof$
\end{assumption}
The local $\frac12$-H\"older continuity in Assumption \ref{a:holder} can be used to upgrade the unique weak solution of \eqref{dX22intro} to a pathwise unique strong solution. To see this, the weaker Engelbert and Schmidt conditions in Assumption \ref{a:speed} produces a global weak solution (unique in law). Assumption \ref{a:holder} allows us to use Yamada-Watanabe's theorem to prove strong uniqueness for  $t\in [0,T_n]$ for a reducing sequence of stopping times $(T_n)_{n\in\N}$. Because $T_n \uparrow \infty$ as $n\uparrow \infty$, this gives global strong uniqueness; hence, global strong existence follows.\footnote{When $\sigma$ is globally $\frac12$-H\"older continuous, the unique strong solution $X_t^x$  of \eqref{dX22intro}  is a martingale. This is because $\sup_{\xi\in\R}\frac{|\sigma(\xi)-\sigma(0)|}{\sqrt{|\xi|}} <\infty$ implies $\int_0^\infty\frac{\xi}{\sigma(\xi)^2}d\xi=\int_{-\infty}^0 \frac{|\xi|}{\sigma(\xi)^2}d\xi=\infty$. Theorem 1 in \cite{Kotani2006} gives the martingale property.
}

\begin{theorem}\label{thm_Main2} Suppose Assumption \ref{a:holder}  holds.  Then, for continuous data $H:\R\to\R$ satisfying \eqref{Hbound1}, we have:
\begin{itemize}

\item[(i)]  If $\int_0^\infty \xi m(d\xi) <\infty$ and $\int_{-\infty}^0 |\xi| m(d\xi)=\infty$, the function $h^*$  in \eqref{def_h} is the unique classical solution $h\in \sC^{1,2}$ of \eqref{CauchyI1} in $\mathcal{H}_{\lambda\uparrow}$.

\item[(ii)]  If $\int_0 ^\infty\xi m(d\xi)=\infty$ and $\int_{-\infty} ^0|\xi| m(d\xi) <\infty$, the function $h^*$  in \eqref{def_h} is the unique classical solution $h\in \sC^{1,2}$ of \eqref{CauchyI1} in $\mathcal{H}_{\lambda\downarrow}$

\item[(iii)]  If $\int_0^\infty \xi m(d\xi) <\infty$ and  $\int_{-\infty} ^0|\xi| m(d\xi) <\infty$, the function $h^*$  in \eqref{def_h} is the unique classical solution $h\in \sC^{1,2}$ of \eqref{CauchyI1} in $\mathcal{H}_{\lambda\uparrow}\cap \mathcal{H}_{\lambda\downarrow}$.
\end{itemize}
\end{theorem}
}
\noindent{\emph{Proof of (i):} {\bf Step 1/2:} In this first step, we consider $H$ positive; that is, $H:\R\to[0,\infty)$. First, we prove that $h^*$ defined in \eqref{def_h} is a classical solution of \eqref{CauchyI1} bounded by $Ke^{\lambda t}\Phi_\lambda(x)$ and satisfies  \eqref{UC2}. 
 
Because $H$ satisfies \eqref{Hbound1}, we can find two  positive constants $(r_0,r)$ such that $H(x)\le r_0+r\Phi_\lambda(x)$ for all $x\in\R$. Therefore, we have the upper bound
\begin{align}\label{Hbound2}
\begin{split}
\E^x[H(X_t)] &\le r_0 + r\E^x[\Phi_\lambda(X_t)]\\
& \le r_0 + r e^{\lambda t} \Phi_\lambda(x),
\end{split}
 \end{align}
where the last inequality follows from $\big(e^{-\lambda t}\Phi_\lambda(X_t)\big)_{t\ge0}$ being a $\P^x$-supermartingale. Because $\Phi_\lambda(x)>0$  is uniformly bounded away from zero, the second inequality \eqref{Hbound2}  ensures that the function { $\tilde h$ in \eqref{tildeh}
satisfies the bound (uniform in $x$)
\begin{align}\label{tildeh2e}
\begin{split}
\tilde{h}(t,x)&\le \frac{r_0}{\Phi_\lambda(x)}+ r e^{\lambda t}\le \frac{r_0}{\inf_{\xi\in \R}\Phi_\lambda(\xi)}+ r e^{\lambda t},\quad t\ge0,\quad x\in\R.
\end{split}
\end{align}
}

 The property $\int_0^\infty \xi m(d\xi) <\infty$ and Theorem \ref{lem_general} ensure that $\varphi_{\lambda \uparrow}(\xi)$ satisfies 
\begin{align}\label{c0}
c_0:= \limsup_{y\uparrow \infty} \frac{\varphi_{\lambda \uparrow}(y)}{y}\in (0,\infty).
\end{align}
Then, for a non-increasing sequence $(s_n)_{n\in\N} \subset (0,\infty)$ with a positive and finite limit $s_\infty:=\lim_{n\uparrow \infty}s_n \in (0,\infty)$, the limit \eqref{finfty2} in Lemma  \ref{cetin2018}(i) gives us the limit in \eqref{UC2} because
\begin{align}\label{UC1}
\begin{split}
\lim_{n\uparrow \infty} \frac{h(s_n,n)}{n}&\le \lim_{n\uparrow \infty} \frac{r_0+ r\E^n[\Phi_\lambda (X_{s_n})]}{ \varphi_{\lambda \uparrow}(n)}\frac{ \varphi_{\lambda \uparrow}(n)}{n}\\
&\le \lim_{n\uparrow \infty} \frac{r_0+ r e^{\lambda s_n} \pld (n) +r\E^n[\plu (X_{s_n})]}{ \varphi_{\lambda \uparrow}(n)}\frac{ \varphi_{\lambda \uparrow}(n)}{n}\\
&\le c_0\lim_{n\uparrow\infty} \frac{r_0' + r\E^n[\varphi_{\lambda\uparrow} (X_{s_n})]}{ \varphi_{\lambda \uparrow}(n)}\\
&=0,
\end{split}
\end{align}
where the  constant  $c_0$ is from \eqref{c0} and $r'_0$ is some positive irrelevant constant.

To see that the PDE \eqref{CauchyI1} holds, we change coordinates. To this end, first assume that $H$ satisfies
\begin{align}\label{Hbound11}
H(\xi) \le r_0 + r\varphi_{\lambda \uparrow}(\xi),\quad \xi \in\R,
\end{align}
for positive constants $(r_0,r)$. { Due to the continuity of $\sigma$, \eqref{SLODE}  yields that  $\varphi_{\lambda\uparrow}$ is a strictly increasing and strictly convex function with $\varphi_{\lambda\uparrow},\varphi^{-1}_{\lambda \uparrow}\in \mathcal{C}^2$.} The continuous function 
\begin{align}\label{F1}
F(y) := H\big(\varphi_{\lambda\uparrow}^{-1} (y)\big),\quad 
y>\underline{y}:= \lim_{\xi\downarrow -\infty} \varphi_{\lambda\uparrow}(\xi),
\end{align}
is of at most linear growth and satisfies $\lim_{y\downarrow \underline{y}}F(y)<\infty$.  Furthermore, for a fixed constant $T\in(0,\infty)$, we define the process 
\begin{align}\label{Y33}
Y_t:= e^{\lambda (T-t)} \varphi_{\lambda\uparrow}(X_t),\quad t\in[0,T], \quad \lambda>0,
\end{align}
with the local martingale dynamics
\begin{align}\label{dY33}
\begin{split}
dY_t&= e^{\lambda (T-t)} \varphi'_{\lambda\uparrow}(X_t)\sigma(X_t)dB_t\\
&=  e^{\lambda (T-t)} \varphi_{\lambda \uparrow}'\big(\varphi^{-1}_{\lambda \uparrow}(e^{-\lambda (T-t)} Y_t)\big)\sigma\big( \varphi^{-1}_{\lambda \uparrow}(e^{-\lambda (T-t)}Y_t)\big) dB_t\\
&= \alpha(t,Y_t)dB_t,
\end{split}
\end{align}
where we have defined the volatility function
\begin{align}\label{alphaH}
\begin{split}
\alpha(t,y) & := e^{\lambda (T-t)} \varphi_{\lambda \uparrow}'\big( \varphi^{-1}_{\lambda \uparrow}(e^{-\lambda (T-t)}y)\big)\sigma\big(\varphi^{-1}_{\lambda \uparrow}(e^{-\lambda (T-t)} y)\big),\quad 
y>\underline{y}, \quad t\ge0.
\end{split}
\end{align}
 {Because $\varphi_{\lambda\uparrow},\varphi^{-1}_{\lambda \uparrow}\in \mathcal{C}^2$, $\alpha(t,y)$ is continuous in $(t,y)$ and locally H\"older continuous in $y$ with exponent $\frac12$. Therefore, }Theorem 3.2 in Ekstr\"om and Tysk \cite{ET2009} guarantees that  the function
\begin{align}
f(t,y) := \bar{\E}^{t,y}[F(Y_T)],\quad t\in[0,T],\quad y>\underline{y},
\end{align}
where $\bar{\E}^{t,y}$ denotes the expectation with respect to the law of $(Y_u)_{u\in [t,T]}$ conditional on $Y_t=y$, is a classical solution of the Cauchy problem
\begin{align}\label{PDE:main}
\begin{cases}
& f(T,y) = F(y),\quad y>\underline{y},\\
&0= f_t(t,y) + \frac12 \alpha(t,y)^2f_{yy}(t,y),\quad y>\underline{y}, \quad t\in(0,T).
 \end{cases}
\end{align}
Then, the function $h^*$ from \eqref{def_h} satisfies the PDE \eqref{CauchyI1} because $\varphi_{\lambda\uparrow}$ solves the Sturm-Liouville ODE \eqref{SLODE} and we have the relation
\begin{align}
\begin{split}
f\big(t,y\big) &= \bar{\E}^{t,y}[F(Y_T)]\\
&= \E^{y^\circ(t)}[H(X_{T-t})]\\
&= h^*\big(T-t,\varphi_{\lambda\uparrow}^{-1}(e^{-\lambda(T-t)}y)\big),
\end{split}
\end{align}
where we have defined $y^\circ(t):=\varphi_{\lambda\uparrow}^{-1}(e^{-\lambda(T-t)}y)$ for $t\in[0,T]$ and $y>\underline{y}$.

Second, a similar argument but replacing \eqref{Hbound11} with
\begin{align}\label{Hbound3}
H(\xi) \le r_0 + r\varphi_{\lambda \downarrow}(\xi),\quad \xi \in\R,
\end{align}
and replacing \eqref{Y33} with $Y_t:= e^{\lambda (T-t)} \varphi_{\lambda\downarrow}(X_t)$ when changing coordinates, shows that  $h$ from \eqref{def_h} satisfies the PDE \eqref{CauchyI1} again. 

Third, by writing
\begin{align}\label{Hbound2222a}
H(\xi) = H^1(\xi) + H^2(\xi) - H(0),\quad H^1(\xi) := H(\xi \vee 0),\quad  H^2(\xi) := H(\xi \land 0), 
\end{align}
and noting that when $H$ satisfies  \eqref{Hbound1}, $H^1$ satisfies \eqref{Hbound11} and $H^2$ satisfies \eqref{Hbound3}.  Then, for $i\in\{1,2\}$, the functions
\begin{align}\label{def_hi}
\begin{split}
h^i(t,x):&=\E^x[H^i(X_t)],\quad t\ge0,\quad x\in\R,
\end{split}
\end{align}
satisfy the PDEs
\begin{align}\label{Cauchy12}
\begin{cases}
&h^i(0,x) = H^i(x),\quad x\in \R,\\
&h^i_t(t,x)  = \frac12\sigma(x)^2h^i_{xx}(t,x),\quad t>0,\quad x\in\R.
\end{cases}
\end{align}
Therefore, $h(t,x):= h^1(t,x)+h^2(t,x)-H(0)$ is the function in \eqref{def_h} and by using  the PDEs in \eqref{Cauchy12}, we see that $h$ satisfies \eqref{CauchyI1}.\ \\

\noindent {\bf Step 2/2:} We consider $H:\R\to\R$ and write $H(\xi)=H^+(\xi)-H^-(\xi)$ where $H^+,H^-:\R\to[0,\infty)$ are defined by $H^+(\xi):= H(\xi)\vee 0$ and $H^-(\xi):= -\big( H(\xi)\land 0\big)$ for $\xi \in \R$. The first step ensures that the functions
\begin{align}\label{def_hpm}
\begin{split}
h^\pm(t,x):&=\E^x[H^\pm(X_t)],\quad t\ge0,\quad x\in\R,
\end{split}
\end{align}
satisfy (uniquely)
\begin{align}\label{CauchyI1pm}
\begin{cases}
&h^\pm(0,x) = H^\pm(x),\quad x\in \R,\\
&h^\pm_t(t,x)  = \frac12\sigma(x)^2h^\pm_{xx}(t,x),\quad t\ge0,\quad x\in\R,
\end{cases}
\end{align}
as well as the limit in \eqref{UC2}:
\begin{align}\label{UC2prime}
\lim_{n\uparrow \infty} \frac{h^\pm(s_n,n)}{n}=0,\quad \text{whenever}\quad \infty > s_n \ge s_{n+1} \text{ and }s_\infty:=\lim_{n\uparrow \infty} s_n >0.
\end{align}
Then, the difference $h:= h^+-h^-$ is the function in \eqref{def_h} and by taking differences in \eqref{CauchyI1pm}, we see that $h$ satisfies the PDE \eqref{CauchyI1} and the limit in \eqref{UC2}.

\ \\
\noindent (ii) and (iii): These are similar to (i) and are omitted. 

$\endproof$

Based on Theorem \ref{thm_Main2}, the value function $h^*(t,x)$ in \eqref{def_h} can exhibit a boundary layer at $t=0$ in the following sense: Consider the mean $H(\xi):=\xi$, $\xi \in \R$,  which satisfies \eqref{Hlinear}. Then, whenever $\int_0^\infty \xi m(d\xi) <\infty$, the value function $h^*(t,x)$ in \eqref{def_h}  satisfies \eqref{intro_FL} as $x\uparrow\infty$ because Theorem \ref{thm_Main2}(i) gives  us
\begin{align}\label{FL1}
\begin{split}
\lim_{x\uparrow\infty}\lim_{t\downarrow 0}\frac{h^*(t,x)}{x} &= 1,\\
\lim_{t\downarrow 0}\lim_{x\uparrow\infty}\frac{h^*(t,x)}{x} &=0.
\end{split}
\end{align}
Similarly, whenever $\int_{-\infty} ^0|\xi| m(d\xi) <\infty$, the value function $h^*(t,x)$ in \eqref{def_h} satisfies  \eqref{intro_FL}  as $x\downarrow-\infty$ because Theorem \ref{thm_Main2}(ii) gives  us
\begin{align}\label{FL2}
\begin{split}
\lim_{x\downarrow -\infty}\lim_{t\downarrow 0}\frac{h^*(t,x)}{x} &= 1,\\
\lim_{t\downarrow 0}\lim_{x\downarrow-\infty}\frac{h^*(t,x)}{x} &=0.
\end{split}
\end{align}

\subsection{Examples}

The first example covers a class of real valued strict local martingales frequently used in  finance (see, e.g.,  Z\"uhlsdorff \cite{Zuhl2001}  and  Andersen \cite{Andersen2011}).
\begin{example}\label{ex:QNV} Quadratic normal  volatility models use dynamics defined by
\begin{align}\label{BC1998}
dX_t := \big(\alpha_0 + \alpha_1X_t + \alpha_2X_t^2\big)dB_t,\quad X_0 \in\R,
\end{align}
and have been widely used in financial economics (see Carr, Fisher, and Ruf \cite{CFR2013} for an overview). Depending on the root configuration $(\alpha_0+\alpha_1\xi+\alpha_2\xi^2=0, \,\xi\in\R)$ relative to the initial value $X_0$, the solution to the SDE \eqref{BC1998} is bounded or unbounded from above and/or below. For example, in a Radner equilibrium model with limited stock-market participation, the following SDE is endogenously derived in Eq. (27) in Basak and Cuoco \cite{BC1998}: 
\begin{align}\label{BC1998a}
dX_t = -X_t(1+X_t)\sigma dB_t,\quad X_0 >0,
\end{align}
for a constant $\sigma\in(0,\infty)$. The dynamics \eqref{BC1998a} produce a nonnegative strict local martingale. Another specification of \eqref{BC1998} is the no-real-root specification  used for option pricing in Section 3.6 in Z\"uhlsdorff \cite{Zuhl2001}  and Eq. (4.1) in Andersen \cite{Andersen2011}. This process is exogenously  given by the dynamics
\begin{align}\label{BC1998b}
dX_t = b\Big(1+\big(\frac{X_t-a}b\big)^2\Big)dB_t,\quad X_0 \in\R,
\end{align}
for constants $(a,b)$ with $b\in(0,\infty)$. The dynamics \eqref{BC1998b} produce a real-valued strict local martingale.  Because
\begin{align}
\int_0 ^\infty \frac{\xi }{\big(b^2+(a-\xi)^2\big)^2}d\xi<\infty,\quad \int_{-\infty}^0 \frac{|\xi| }{\big(b^2+(a-\xi)^2\big)^2}d\xi<\infty,
\end{align}
we see from Theorem \ref{thm_Main}(iii) that $h(t,x)$ in \eqref{def_h} vanishes as $x\to \pm\infty$ for $t>0$. In this case, the mean function $H(\xi) := \xi$, $\xi\in\R$, produces a double boundary layer in the sense that for $t>0$ we have the limits in both \eqref{FL1} and \eqref{FL2}.
$\endproof$
\end{example}

The second example is based on the two dimensional Bessel process.

\begin{example}[Continuation of Example \ref{bes2D}] Let $(X_t)_{t\ge0}$ be the logarithm of the two dimensional dimensional Bessel  process \eqref{dX2}. We claim that 
\begin{align}\label{f2D}
\begin{split}
\E^x[X_t] =x + \int_{e^x}^\infty \frac1r e^{- \frac{r^2}{2t}}dr,\quad x\in\R,\quad t\ge0.
\end{split}
\end{align}
To see this, we define the function 
\begin{align}\label{f2DD}
\begin{split}
h(t,x):= x + \int_{e^x}^\infty \frac1r e^{- \frac{r^2}{2t}}dr, \quad x\in\R,\quad t\ge0.
\end{split}
\end{align}
By computing $t$ and $x$ derivatives in \eqref{f2DD}, we see that the PDE in  \eqref{CauchyI1} holds. Furthermore, for $t>0$,  L'Hopital's rule produces the limit
\begin{align}
\begin{split}
\lim_{x\downarrow -\infty}\frac1x\int_{e^x}^\infty \frac1r e^{- \frac{r^2}{2t}}dr &= -\lim_{x\downarrow -\infty} e^{- \frac{e^{2x}}{2t}}\\&=-1.
\end{split}
\end{align}
Therefore, for $t>0$, the function $h$ in \eqref{f2DD} has the limit in \eqref{UC3}. Because 
\begin{align}
\int_0 ^\infty\xi e^{2\xi}d\xi=\infty,\quad \int_{-\infty} ^0|\xi| e^{2\xi}d\xi<\infty,
\end{align}
we can use the uniqueness part of Theorem \ref{thm_Main}(ii) to see that \eqref{f2D} holds.  Consequently, the boundary layer limits  in \eqref{FL2} hold.

As an aside, the limit in \eqref{UC3} trivially holds because we have
\begin{align}\label{2Dfinitelimit}
\lim_{x\downarrow -\infty} \E^x[X_t] =\frac12 \big( \log(2) + \log(t)-\gamma \big)\in \R,\quad t>0,
\end{align}
where $\gamma$ is the Euler-Mascheroni constant ($\gamma \approx 0.57721$).\footnote{When $X_t$ is an inverse  three dimensional Bessel process (which is positive), Example 2.2.2 in \cite{CH2005} gives a limit similar to \eqref{2Dfinitelimit} with $\lim_{x\uparrow \infty} \E^x[X_t]<\infty$ for $t>0$. } 
$\endproof$
\end{example}

{

\section{Uniqueness for higher moments} 
Under Assumption \ref{a:holder}, we denote by $X_t^x$ the unique strong solution of 
\eqref{dX22intro} for $x\in \R$ and $t\ge0$. This section uses the above results to regularize the strict local martingale $X^x_t$ into a  martingale $N_t$. This regularization allows us to prove uniqueness of classical solutions to an altered PDE when the continuous data $H$ is of at most polynomial growth, i.e.,
\begin{align}\label{Hp}
|H(\xi)| \le c(1+|\xi|^p),\quad \xi \in \R,
\end{align}
where $c\in(0,\infty)$ and $p\in(1,\infty)$ are constants ($c$ and $p$ can vary with $H$). For $T\in(0,\infty)$, we define the martingale
\begin{align}\label{def:Nt}
\begin{split}
N_t :&=\E[X^x_T | \sF_t] \\
&= h^*(T-t,X^x_t),\quad t\in[0,T],  \quad x\in\R,
\end{split}
\end{align}
where $h^*$ is from Theorem \ref{thm_Main2}. We shall show in  Lemma \ref{l:comparison} below that  $x\to h^*(t,x)$ is strictly increasing under the hypothesis of Theorem \ref{thm_Main2}. The difficulty in proving this seemingly trivial result lies in the fact that  $X_t^x$ is a strict local martingale. Indeed, when $X_t^x$ is a martingale, $\E[X^x_t]=x$ is trivially strictly increasing. Moreover, $X_t^x$ being a strict local martingale also implies that we cannot use strict comparison results for SDEs based on Lipschitz continuity like Theorem 33.6 in Kallenberg (2021) and Theorem in IX.3.8 in Revuz and Yor \cite{RevYor}.  On the other hand, even when $X_t^x$ is a strict local martingale, Theorem 1.4 in Lowther (2008), shows that $x\to h^*(t,x)$ is non-decreasing but this property is insufficient to produce the inverse function  $(h^*)^{-1}(t,\cdot)$ we need below. 

\begin{lemma} \label{l:comparison} Let $F:\R\to\R$ be continuous, satisfy \eqref{Hbound1},  and be strictly increasing. Under Assumption \ref{a:holder}, for each $t\ge0$, the function $\R\ni x\to\E[F(X_t^x)]$  is strictly increasing.
\end{lemma}

\proof Theorem IX.3.8 in \cite{RevYor} or Theorem V.43.1 in \cite{RV2000} ensures that  $x<y$ implies $X _t^x \le X_t^y$, $\P$-a.s., for all $t\ge0$. We claim that for all $x<y$, we have
\begin{align}\label{t0}
t_0 := \inf \{ t\ge0 : \P(X_t^x = X_t^y)=1\}= \infty.
\end{align}
This claim gives us that $\R\in x\to\E[F(X_t^x)]$ is strictly increasing. We argue by contradiction and assume $t_0 \in [0,\infty)$.  

\noindent {\bf Step 1/3:} If $t_0=0$, we can find $t_n\downarrow 0$ such that $\P(X_{t_n}^x = X_{t_n}^y)=1$ for all $n\in \N$. Then, the set $\Omega^\circ:= \{\omega \in\Omega \;|\; \forall n\in\N:X^x_{t_n}(\omega) = X_{t_n}^y(\omega)\}$ satisfies $\P(\Omega^\circ)=1$, and so path continuity gives the contradiction 
\begin{align*}
 0&=\lim_{n\to\infty} \big(X_{t_n}^x(\omega)-X_{t_n}^y(\omega)\big)=x-y <0,\quad \forall  \omega \in \Omega^\circ.
\end{align*}

\noindent {\bf Step 2/3:} If $t_0 \in (0,\infty)$, we claim\footnote{In \eqref{mon1}, we cannot consider  intersections like $(X_\epsilon^x\le x) \cap (X_\epsilon^y\ge y)$ because these sets can be  nullsets for all $\epsilon>0$. For example, $X_t^x := x + B_t$ gives the nullset $(X_\epsilon^x\le x) \cap (X_\epsilon^y\ge y)=(B_\epsilon =0)$.}
\begin{align}
\exists \epsilon \in(0,t_0):\;\P( X_\epsilon^x\le \tfrac{x+y}2, X_\epsilon^y\ge y)>0 \text{ and }\P( X_\epsilon^x\le x, X_\epsilon^y\ge \tfrac{x+y}2)>0. \label{mon1}
\end{align}
We argue by contradiction and assume \eqref{mon1} fails; that is,  we assume
\begin{align}\label{mon1a0}
\forall \epsilon \in(0,t_0): \;\P( X_\epsilon^x\le \tfrac{x+y}2, X_\epsilon^y\ge y)=0 \text{ or }\P( X_\epsilon^x\le x, X_\epsilon^y\ge \tfrac{x+y}2)=0.
\end{align}
Equivalently, we assume
\begin{align}\label{mon1a1}
\forall \epsilon \in(0,t_0): \;\P(X_\epsilon^x> \tfrac{x+y}2\;\text{or} \;X_\epsilon^y< y)=1\text{ or }\P( X_\epsilon^x> x\;\text{or} \; X_\epsilon^y< \tfrac{x+y}2)=1.
\end{align}
Define the sequence $\epsilon_n:=\frac1n$ for $n\in\N$ big enough such that $\epsilon_n \in (0,t_0)$. Then, based on \eqref{mon1a1}, there exists a subsequence $(\epsilon'_n)_{n\in\N}\subset (\epsilon_n)_{n\in\N}$ such that at least one of the following two statements holds 
\begin{align}
&\forall n\in\N:\P(X_{\epsilon'_n}^x> \tfrac{x+y}2\;\text{or} \;X_{\epsilon'_n}^y< y)=1, \\
&\forall n\in\N:\P( X_{\epsilon_n'}^x> x\;\text{or} \; X_{\epsilon'_n}^y< \tfrac{x+y}2)=1.\label{mon1a11}
\end{align}
Say that \eqref{mon1a11} holds (the argument in the other case is similar).
We define
$$
\Omega':= \{ \omega\in \Omega\,|\, \forall n\in\N:X_{\epsilon'_n}^x(\omega)> x \;\text{or} \;X_{\epsilon'_n}^y(\omega)< \tfrac{x+y}2 \}.
$$
Eq.  \eqref{mon1a11} gives $\P(\Omega')=1$. Because $\tfrac{x+y}2<y$, path continuity gives the set inclusion
\begin{align}
\begin{split}
\label{mon1a11bb} \
\Omega' &\subseteq  \{ \omega\in \Omega\,|\, \exists N(\omega)\in\N:  \forall n\ge N(\omega), \, X_{\epsilon'_n}^x(\omega)>x \}=:\Omega''.
\end{split}
\end{align}
To show the contradiction $\P(\Omega'')=0$, we define the sets 
\begin{align}\label{En}
E_n^\delta:=  \{ \omega\in \Omega\,|\, X_{\epsilon'_n}^x(\omega)\ge x+\delta\},\quad n\in \N,\quad \delta >0.
\end{align}
Because the sets
\begin{align}\label{Enn}
\Omega'''(\delta) :=  \{ \omega\in \Omega\,|\,  \exists N(\omega)\in\N: \forall n\ge N(\omega),  X_{\epsilon'_n}^x(\omega)\ge x+\delta \},\quad \delta >0,
\end{align}
satisfy $\Omega'''(\delta) \subseteq (E^\delta_n \text{ i.o.})$, $\Omega'''(\delta)\subseteq \Omega'''(\delta')$ for $\delta> \delta'>0$, and  $\cup_{\delta>0}\Omega'''(\delta)= \Omega''$, it suffices to prove $\P (E^\delta_n \text{ i.o.})=0$ for each $\delta >0$ to justify  $\P(\Omega'')=0$. By the Borel-Cantelli lemma, $\P (E^\delta_n \text{ i.o.})=0$ is ensured by $\sum^\infty \P(E_n^\delta)<\infty$. To this end, the hitting times $T_{x+\delta}:=\inf\{t>0:X^x_t=x+\delta\}$ for $x\in\R$ and $\delta >0$ satisfy
\begin{align}\label{uc1}
\P(E_n^\delta) \le \P(T_{x+\delta}\le \epsilon'_n) \le e \E[e^{-T_{x+\delta}/\epsilon'_n}],\quad n\in\N,
\end{align}
where the last inequality follows from 
$$
\E[e^{-T_{x+\delta}/\epsilon'_n}] \ge \E[e^{-T_{x+\delta}/\epsilon'_n}1_{T_{x+\delta}\le \epsilon'_n}] \ge e^{-1} \P(T_{x+\delta}\le \epsilon'_n).
$$
The proposition on p. 258 in Kotani and Watanabe (1982) gives us the limit
\begin{align}
\lim_{n\to\infty} \sqrt{\epsilon_n'}\big(-\log(\E[e^{-T_{x+\delta}/\epsilon'_n}])\big) = \sqrt{2}\int_x^{x+\delta} |\sigma(\xi)|d\xi\in(0,\infty).
\end{align}
Consequently, there exists a constant $C^\delta>0$ (independent of $n\in\N$) such that for large $n\in \N$, we have $-\log(\E[e^{-T_{x+\delta}/\epsilon'_n}] )\ge C^\delta/ \sqrt{\epsilon_n'}$. Therefore, \eqref{uc1} gives
$$
\sum^\infty \P(E_n^\delta) \le e \sum^\infty\E[e^{-T_{x+\delta}/\epsilon'_n}]\le e \sum^\infty e^{- C^\delta/ \sqrt{\epsilon_n'}}\le e \sum^\infty e^{- C^\delta \sqrt{n}}<\infty.
$$

\noindent {\bf Step 3/3:} For $\epsilon$ as in \eqref{mon1}, the Markov property gives
\begin{align}\label{mon3}
\begin{split}
f(X^x_\epsilon) = \E[X^x_{t_0}|\sF_\epsilon],\quad f(z):= \E[X^z_{t_0-\epsilon}],\quad z\in \R.
\end{split}
\end{align}
Because $t_0-\epsilon <t_0$, the definition of $t_0$ in \eqref{t0} gives $\P(X_{t_0-\epsilon}^x = X^y_{t_0-\epsilon})<1$ which combined with $\P(X_{t_0-\epsilon}^x\le X^y_{t_0-\epsilon})=1$ gives 
\begin{align}\label{mon4}
\begin{split}
f(x)= \E[X^x_{t_0-\epsilon}]< \E[X^y_{t_0-\epsilon}]=f(y).
\end{split}
\end{align}
We split $[x,y]$ into $[x,\frac{x+y}2]$ and $[\frac{x+y}2,y]$. The strict inequality in \eqref{mon4} gives us that at least of the following two inequalities hold
\begin{align}\label{mon4e}
\begin{split}
f(x)< f(\tfrac{x+y}2),\quad f(\tfrac{x+y}2)<f(y).
\end{split}
\end{align}
Both cases in \eqref{mon4e} are similar and it suffices to consider $f(\tfrac{x+y}2)<f(y)$. Then,
\begin{align}\label{mon4a}
\begin{split}
f\big(X^x_\epsilon(\omega)\big) \le f(\tfrac{x+y}2) < f(y) \le f\big(X^y_\epsilon(\omega)\big),\quad \omega \in (X_\epsilon^x\le \tfrac{x+y}2)\cap(X_\epsilon^y\ge y). 
\end{split}
\end{align}
By taking expectations in \eqref{mon4a} and using \eqref{mon1} and $\P(X_\epsilon^x\le X^y_\epsilon)=1$,  we get
\begin{align}\label{mon4aa}
\begin{split}
\E[f(X^x_\epsilon)] < \E[f(X^y_\epsilon)].
\end{split}
\end{align}
On the other hand, we can find $t_n\downarrow t_0$ such that $\P(X_{t_n}^x = X_{t_n}^y)=1$ for all $n\in \N$. Therefore, path continuity gives  $\P(X_{t_0}^x = X_{t_0}^y)=1$. Consequently, the Markov property in \eqref{mon3} produces a contradiction with \eqref{mon4aa}:
\begin{align}\label{mon5}
\begin{split}
\E[f(X^x_\epsilon)] = \E[X^x_{t_0}]= \E[X^y_{t_0}]=\E[f(X^y_\epsilon)].
\end{split}
\end{align}
$\endproof$

Let $h^*$ be as in \eqref{def:Nt} and denote its inverse by $(h^*)^{-1}(t,\cdot): \big(\underline{y}(t),\overline{y}(t)\big) \to \R$ where
$$
\underline{y}(t):= \lim_{x\downarrow -\infty}h^*(t,x)\in[-\infty,\infty),\quad \overline{y}(t):= \lim_{x\uparrow \infty}h^*(t,x)\in (-\infty,\infty],\quad t\in[0,T].
$$

The martingale $N_t$ in \eqref{def:Nt} has the Markovian dynamics 
\begin{align}
\begin{split}
dN_t &= h^*_x(T-t,X^x_t) \sigma(X^x_t)dB_t \\
&= \tilde\sigma(T-t,N_t)dB_t,
\end{split}
\end{align}
where the time-dependent volatility function $\tilde\sigma$ is defined as
\begin{align}
\begin{split}
\tilde\sigma(u,y):= h^*_x\big(u,(h^*)^{-1}(u,y)\big) \sigma\big((h^*)^{-1}(u,y)\big),\quad y \in\big(\underline{y}(u),\overline{y}(u)\big),\quad u \in [0,T].
\end{split}
\end{align}

The proof of the next uniqueness result uses Doob's maximal martingale inequality. Doob's inequality  can fail for strict local martingales.\footnote{For example, $Y_t$ in \eqref{dX3d} has $\E[Y^2_t]<\infty$ and $\E[\sup_{s\in[0,t]}Y_s^2]=\infty$ for $t>0$.}

\begin{theorem}\label{thm_Main3} In the setting of Theorem \ref{thm_Main2}, suppose $\E[|X^x_T|^p]<\infty$ for all $x\in\R$ and fixed $T\in(0,\infty)$ and $p>1$. Then, for continuous data $H:\R\to\R$ satisfying \eqref{Hp}, there is at most one classical solution $g\in \sC^{1,2}$ of
\begin{align}\label{CauchyI11}
\begin{cases}
&g(0,y) = H(y),\quad y\in \R,\\
&g_t(u,y)= \frac12\tilde\sigma(u,y)^2 g_{yy}(u,y),\quad y\in\big(\underline{y}(u),\overline{y}(u)\big)\quad u\in(0,T],
\end{cases}
\end{align}
satisfying $|g(u,y)| \le c_0(1+|y|^p)$ for all $u\in[0,T]$ and $y\in\big(\underline{y}(u),\overline{y}(u)\big)$ for a constant $c_0$ ($c_0$ can vary with $g$).
\end{theorem}

\proof  Let $g$ be as in the statement. It\^o's lemma and the PDE in \eqref{CauchyI11} produce the local martingale dynamics
$$
dg(T-t,N_t) = g_y(T-t,N_t)\tilde\sigma(T-t,N_t)dB_t,\quad t\in[0,T].
$$
For $n\in\N$ with $(h^*)^{-1}(T,y)\in (-n,n)$, we define the passage times $T_n :=\inf\{t>0 : |N_t|\ge n\}$. Because $g(T-t\land T_n,N_{t\land T_n})$, $t\in[0,T]$ and $n\in\N$, is a (bounded) martingale, we have
\begin{align}\label{keyy1}
g(T,y) = \E^{(h^*)^{-1}(T,y)}[g(T-T\land T_n,N_{T\land T_n})],\quad n\in\N,\quad  y\in\big(\underline{y}(T),\overline{y}(T)\big).
\end{align}
The proof is concluded by justifying that we can pass $n\to \infty$ inside the expectation in \eqref{keyy1} to produce the representation
\begin{align}\label{keyy1}
g(T,y) = \E^{(h^*)^{-1}(T,y)}[H(N_T)],\quad  y\in\big(\underline{y}(T),\overline{y}(T)\big).
\end{align}
Dominated convergence can be used in \eqref{keyy1} because 
\begin{align}
\begin{split}
\E^{(h^*)^{-1}(T,y)}\big[\sup_{u\in[0,T]}|g(T-u,N_u)|\big] &\le c_0\big(1+\E^{(h^*)^{-1}(T,y)}\big[\sup_{u\in[0,T]}|N_u|^p\big]\big)\\
&\le c_0\left(1+\Big(\frac{p}{p-1}\Big)^p\E^{(h^*)^{-1}(T,y)}\big[|N_T|^p\big]\right)\\
&=c_0\left(1+\Big(\frac{p}{p-1}\Big)^p\E^{(h^*)^{-1}(T,y)}\big[|X_T|^p\big]\right),
\end{split}
\end{align}
which is finite by assumption. The second inequality uses Doob's maximal inequality applied to the submartingale $|N_t|$.
$\endproof$
\ \\

When $\sigma$ is Lipschitz, the martingale $X^x_t$ has all moments. We leave it open to find conditions on $\sigma$ ensuring $X^x_t$ is $p$ integrable when $\sigma$ is only locally $\frac12$-H\"older continuous. %To illustrate the subtlety of this integrability requirement, just consider the strict local martingale $Y_t>0$ in \eqref{dX3d}. This process has $\E[Y^2_t]<\infty$ but $\E[Y^4_t]=\infty$.

\appendix
\section{Properties of $\lambda$-harmonic functions}\label{appA}
This appendix proves properties of $\lambda$-harmonic functions which we are unable to find references for.
\begin{proposition} \label{p:harmonic}Suppose Assumption \ref{a:speed} holds. Let $\lambda>0$ and $u:\R \to (0,\infty)$ be a $\lambda$-harmonic function for the diffusion in \eqref{dX22intro}. Then, we have:
	\begin{enumerate}
		\item[(i)] $u$ is strictly convex.  
		\item[(ii)] $\lim_{t\downarrow 0}\E^{x}[e^{-\lambda t}u(X_t)]=u(x)$, $x\in\R$, and $u$ is $\lambda$-excessive.
	\end{enumerate}
\end{proposition}
\proof
(i):  Let $x,y \in \R$ with $x<y$ and $a\in (0,1)$,   and recall  the passage times defined in \eqref{Ty}. The optional stopping  theorem produces
	\begin{align}\label{convex1}
		\begin{split}
&			u(ax + (1-a)y) \\
&=\E^{ax + (1-a)y}[e^{-\lambda  T_x\land T_y}u(X_{ T_x\land T_y})]\\
			&<  \E^{ax + (1-a)y}[u(X_{T_x\land T_y})]\\
			& = u(x)\P^{ax + (1-a)y}(X_{T_x\land T_y} = x) + u(y)\P^{ax + (1-a)y}(X_{T_x\land T_y} = y),
		\end{split}
	\end{align}
where the strict inequality follows from  $\P^{ax + (1-a)y}(T_x\wedge T_y>0)=1$ and
\[
\E^{ax + (1-a)y}[(1-e^{-\lambda  T_x\land T_y})u(X_{ T_x\land T_y})]\geq \min\{u(x),u(y)\}\E^{ax + (1-a)y}[1-e^{-\lambda  T_x\land T_y}]>0.
\]
	Because $X_t$ in \eqref{dX22intro} is on the natural scale, we have $\P^{ax + (1-a)y}(X_{T_x\land T_y} = x) = a$ and $u$'s strict convexity follows from \eqref{convex1}.

(ii): Let $x\in \R$ be arbitrary and $n\in \N$ large enough so that $|x|<n$. For $T_n$ defined in \eqref{Ty} we have $\P^x(0<T_n<\infty)=1$ and $e^{-\lambda t}u(X_t)$,  $t\geq 0$, is a $\P^x$-supermartingale. Then, we have
	\begin{align*}
\begin{split}
			\lim_{t\downarrow 0}\E^{x}[e^{-\lambda t}u(X_t)]&=\lim_{t\downarrow 0}\E^{x}[e^{-\lambda t}u(X_t)1_{t< T_n}]+\lim_{t\downarrow 0}\E^{x}[e^{-\lambda t}u(X_t)1_{t\geq  T_n}]\\
			&=\E^{x}[\lim_{t\downarrow 0}u(X_t)]+\lim_{t\downarrow 0}\E^{x}[e^{-\lambda t}u(X_t)1_{t\geq  T_n}]\\
			&=u(x)+\lim_{t\downarrow 0}\E^{x}[e^{-\lambda t}u(X_t)1_{t\geq  T_n}],
\end{split}
	\end{align*}
where the last equality is due to the dominated convergence theorem. The last limit is zero because the supermartingale property of $e^{-\lambda t}u(X_t)$, $t\geq 0$, gives
\begin{align*}\begin{split}
	\lim_{t\downarrow 0}\E^{x}[e^{-\lambda t}u(X_t)1_{t\geq  T_n}]&\leq \lim_{t\downarrow 0}\E^{x}[e^{-\lambda T_n}u(X_{T_n})1_{t\geq  T_n}]\\
	&=\E^{x}[e^{-\lambda T_n}u(X_{T_n}) \lim_{t\downarrow 0}1_{t\geq  T_n}]=0,
\end{split}
\end{align*}
where the first equality uses the dominated convergence theorem and \\$\E^{x}[e^{-\lambda T_n}u(X_{T_n})]\leq u(x)<\infty$ and the second equality uses  $\P^x(0<T_n)=1$.
$\endproof$

}
{
\section{On weak solutions} \label{s:weak}
This Appendix shows that the definition of a weak solution we use coincides with Sawyer \cite{Sawyer}.\footnote{   Sawyer \cite{Sawyer} considers locally bounded $h$ (possibly non-continuous). However, to deal with our initial condition $h(0,x) = H(x)$, we restrict Sawyer's definition to $h\in \sC$.} Consider a finite interval $[a,b]\subset \R$ with  $a<b$  and  $f\in \mathcal{C}^{2}([a,b])$, where the existence and  continuity of  derivatives at the boundaries are only required to exist from the interior. If  $h\in \mathcal{C}^{1,2}$ is a classical solution of \eqref{CauchyI1}, then for $t>0$ we have
\begin{align*}
\int_a^b h_t(t,x) \frac{2}{\sigma^2(x)}f(x)dx &=\int_a^b h_{xx}(t,x)f(x)dx\\
&=\big\{f(x)h_x(t,x)-f'(x)h(t,x)\big\}\big|_a^b+\int_a^b h(t,x)f''(x)dx,
\end{align*}
by integrating by parts twice. 

\begin{definition}[Weak solution]\label{def_weakPDE} $h \in \sC$ is a weak solution of \eqref{CauchyI1} if (i)  $h(0,x)=H(x)$ for all $x\in \R$ and (ii) $h$ satisfies
\begin{equation}\label{e:defweak}
	\begin{split} 
\int_a^b \big(h(t_1,x)-h(t_0,x)\big)f(x)m(dx)= \int_{t_0}^{t_1} dt \int_a^b h(t,x)f''(x)dx\\ +f'(a)\int_{t_0}^{t_1}h(t,a)dt -f'(b)\int_{t_0}^{t_1}h(t,b)dt,
\end{split}
\end{equation}
 for all $0<t_0<t_1$, all $a<b$, and for all $f\in \mathcal{C}^{2}([a,b])$ with $f(a)=f(b)=0$.
$\endproof$
\end{definition}
Because $h$ is continuous  in  Definition \ref{def_weakPDE}, allowing $t_0=0$ in \eqref{e:defweak} produces an equivalent definition.

\begin{lemma} Let  $h \in \sC$ satisfy $h(0,x)=H(x)$. Then, $h$ is a weak solution of \eqref{CauchyI1} 
if and only if  \eqref{e:weakInt} holds for all $f\in \sC_c^{1,2}$.
\end{lemma}

\proof To see that \eqref{e:defweak} is implied by \eqref{e:weakInt}, we consider a sequence of functions $(g_n)_{n\in\N} \subset \mathcal{C}^1([t_0,t_1])$ with $g_n(t_i)=g'_n(t_i)=0$ for all $n\in \N$ and $i\in\{0,1\}$ such that for any $u\in \sC([t_0,t_1])$ we have
\[
\lim_{n\uparrow \infty}\int_{t_0}^{t_1}u(t)g'_n(t)dt= u(t_0)-u(t_1).
\]
In other words, $g'_n(t)dt$ converges weakly to the signed Dirac measure $\delta_{t_0}-\delta_{t_1}$.
Similarly, we can find another sequence $(f_n)_{n\in\N} \subset \mathcal{C}^{2}([a,b])$ with $f_n(y)=f_n'(y)=f_n''(y)=0$  for all $n\in \N$ and $y\in \{a,b\}$ such that  for any $u\in \sC([a,b])$ we have
\[
\lim_{n\uparrow \infty}\int_{a}^{b}f_n''(x)u(x)dx= \int_{a}^{b}f''(x)u(x)dx+ f'(a) u(a)-f'(b)u(b). 
\]
Since $g_nf_n \in \mathcal{C}_c^{1,2}([0,\infty]\times \R)$, continuity of $h$ gives the claimed implication.

Conversely, reversing the above argument allows us to conclude that \eqref{e:defweak} implies \eqref{e:weakInt} for any function $f=f(t,x)\in \mathcal{D}$, where
\[
\mathcal{D}:=\Big \{\sum_{i=1}^jg_i(t)  f_i(x): j\geq 1, g_i \in  \mathcal{C}^1_c\big((0,\infty)\big) \mbox{ and } f_i\in  \mathcal{C}^2_c(\R)\Big\}
\]
and $ \mathcal{C}^i_c$ is the class of $i$-times continuously differentiable functions with compact support. Since $\mathcal{D}$ is an algebra that separates points, we conclude by Stone-Wierstrass'  theorem and the continuity of $h$ that \eqref{e:weakInt} holds for all $f \in \mathcal{C}_c^{1,2}\big((0,\infty)\times \R\big)$.
$\endproof$

\begin{lemma} Let  $h \in \sC$ satisfy $h(0,x)=H(x)$. Then, $h$ is a weak solution of \eqref{CauchyI1} 
if and only if 
\begin{equation} \label{e:defweak2}
	\begin{split}
	\int_a^b u^{ab}(x,y)\big(h(t_1,y)-h(t_0,y)\big)m(dy)=-\int_{t_0}^{t_1} h(t,x)dt\\
+\frac{b-x}{b-a}\int_{t_0}^{t_1}h(t,a)dt +\frac{x-a}{b-a}\int_{t_0}^{t_1}h(t,b)dt,
	\end{split}
\end{equation}
for all $a<b$, $x\in (a,b)$, and $0<t_0<t_1$. In \eqref{e:defweak2}, $u^{ab}$ is the symmetric potential kernel 
\[
u^{ab}(x,y):=\frac{(x-a)(b-y)}{b-a}, \qquad x\leq y, \quad (x,y)\in (a,b)^2.
\]
\end{lemma}

\proof Denote by $X^{ab}$ the solution of \eqref{dX22intro} killed at the first exit time from $(a,b)$. The symmetric potential kernel associated with $X^{ab}$ is $u^{ab}(x,y)$ in the sense that for any non-negative Borel function $g$ we have (see, e.g., Corollary VII.3.8 in \cite{RevYor})
\[
\E^x\left[\int_0^{T_{ab}}g(X_t)dt\right]=U^{ab}g(x),\quad U^{ab}g(x):=\int_{a}^b u^{ab}(x,y)g(y)m(dy),\quad x\in [a,b].
\]
When $g\in \mathcal{C}([a,b])$, we have $f:=U^{ab}g \in \mathcal{C}^{2}([a,b])$ with $f(a) = f(b)=0$. Moreover,
 \[
 \frac{\sigma^2(x) f''(x)}{2}=-g(x),\quad x\in (a,b).
 \]

The converse also holds: Given $f\in \mathcal{C}^{2}([a,b])$  with $f(a) = f(b)=0$, there exits $g\in\mathcal{C}([a,b])$ such that $f= U^{ab}g$.\footnote{This can be shown by applying Ito's formula and noticing that $\E^x[f(X_{T_{ab}})]=0$.}

Direct computations show that $f= U^{ab}g$ has derivative
\[
f'(x)=-\frac{1}{b-a}\int_a^x(y-a)g(y)m(dy)+\frac{1}{b-a}\int_x^b(b-y)g(y)m(dy).
\]
For $g\in \mathcal{C}([a,b])$, we insert $f=U^{ab}g$ into \eqref{e:defweak} and use the symmetry of $u^{ab}$ to get
\begin{align*}
	\int_a^b \big(U^{ab}h(t_1,x)-U^{ab}h(t_0,x)\big)g(x)m(dx)=-\int_{t_0}^{t_1}\int_a^b h(t,x)g(x)m(dx)\\
	+\int_a^b\frac{b-y}{b-a}g(y)m(dy)\int_{t_0}^{t_1}h(t,a)dt +\int_a^b\frac{y-a}{b-a}g(y)m(dy)\int_{t_0}^{t_1}h(t,b)dt.
	\end{align*}
Since $g\in \mathcal{C}([a,b])$ is arbitrary and $m$ is absolutely continuous with respect to the Lebesgue measure, we deduce that  a continuous function  $h$ satisfies \eqref{e:defweak} if and only if \eqref{e:defweak2} holds for all $x\in (a,b)$.

$\endproof$

Theorem A2 and Remark 3 in \cite{Sawyer} show that $h$ is a weak solution of \eqref{CauchyI1} if and only if $\big(h(T-t, X_t)\big)_{t\in [0,T]}$ is a local martingale for all $T>0$. 

\begin{theorem}[Sawyer \cite{Sawyer}]\label{t:weak}
A function $h\in \mathcal{C}$ is a weak solution of \eqref{CauchyI1} if and only if $\big(h(T-t, X_t)\big)_{t\in [0,T]}$ is a $\P^x$-local martingale for all $T>0$ and $x\in \R$.
\end{theorem}

}
}

\end{document}